\documentclass{IEEEtran}
\usepackage{cite}
\usepackage{xcolor}
\usepackage{array}
\usepackage{booktabs}
\usepackage{multirow}
\usepackage{amsmath,amssymb,amsfonts}
\usepackage{algorithmic}
\usepackage{graphicx}
\usepackage{epstopdf}
\usepackage{textcomp}
\usepackage[colorlinks,
            linkcolor=blue,
            anchorcolor=blue,
            citecolor=blue]{hyperref}
\def\BibTeX{{\rm B\kern-.05em{\sc i\kern-.025em b}\kern-.08em
    T\kern-.1667em\lower.7ex\hbox{E}\kern-.125emX}}
\begin{document}
\title{Multibeam Hybrid Transmitarray Based on Polarization Rotating Metasurface With Reconfigurable Bidirectional Radiation}
\author{Fan Qin, \IEEEmembership{Member, IEEE}, Yifei Liu, \IEEEmembership{Student Member, IEEE}, Chao Gu, \IEEEmembership{Member, IEEE}, 

Linfeng Zeng, \IEEEmembership{Student Member, IEEE}, Wenchi Cheng, \IEEEmembership{Senior Member, IEEE}, 

Hailin Zhang, \IEEEmembership{Member, IEEE}, and Steven Gao, \IEEEmembership{Fellow, IEEE}
\thanks{Manuscript received 16 January 2024; revised 22 June 2024; accepted 14 July 2024. This work was supported by the National Key Research and Development Program under Grant 2023YFE3011502. \textit{(Corresponding author: Fan Qin.)}}
\thanks{Fan Qin, Yifei Liu, Linfeng Zeng, Wenchi Cheng, and Hailin Zhang are with the School of Telecommunications Engineering, Xidian University, Xi'an 710071, China (e-mail: fqin@xidian.edu.cn; louisliu@stu.xidian.edu.cn; lfzeng@stu.xidian.edu.cn; wccheng@xidian.edu.cn; hlzhang@xidian.edu.cn). }
\thanks{Chao Gu is with the ECIT Institute, Queen’s University Belfast, BT3 9DT Belfast, U.K. (e-mail: chao.gu@qub.ac.uk).}
\thanks{Steven Gao is with the Department of Electronic Engineering, Chinese University of Hong Kong, Hong Kong (e-mail: scgao@ee.cuhk.edu.hk).}}

\maketitle

\begin{abstract}
This paper proposes a bidirectional multibeam hybrid transmitarray (HTA) employing a transmission polarization-rotating metasurface (TPRM). A novel configuration is introduced to facilitate bidirectional beam scanning by combining the transmitarray (TA) and folded-transmitarray (FTA). To accomplish the reconfiguration of both unidirectional and bidirectional radiation states in the +z, -z, and +/-z directions, a polarization switchable multi-feed array (MFA) is placed at the focal plane between the TA and FTA, radiating x-polarization, y-polarization, and 45-degree oblique polarization waves, respectively. Meanwhile, the proposed antenna can achieve multibeam radiation in the three aforementioned states by switching the polarization of the MFA. To demonstrate the operating principle, a prototype has been designed, simulated, and fabricated. The measured results agree well with the simulated results. The simulated and measured results indicate that the proposed design can generate reconfigurable multibeam in both forward and backward directions, either separately or simultaneously. In the unidirectional states, forward and backward beam scanning is achieved within an angular range of +/-30° and +/-22°, respectively, with peak gains of 23.6 dBi and 23.1 dBi. A simultaneous forward and backward beam scanning of +/-40° and +/-22° is achieved in the hybrid radiation state, with peak gains of 19.4 dBi and 19.3 dBi, respectively. The proposed antenna array design offers several advantages, including bidirectional low-loss beam scanning, a simple structure, low power consumption, and a low profile.
\end{abstract}


\section{Introduction}
\label{sec:introduction}
\IEEEPARstart{W}{ITH} the rapid development of wireless communication, navigation, and radar anti-jamming technologies, increased demands and requirements are imposed on multibeam scanning antennas to establish cost-effective solutions with wide bandwidth and simple feed networks. Beam scanning technology has versatile applications in image processing, remote sensing, and cellular mobile communication [1]–[3]. Due to the characteristics of agile beam steering and fast switching time, electronic beam-switching is gradually replacing traditional mechanical beam steering. Methods for achieving multibeam antennas include phased array antennas, which use controllable elements for beam steering; Butler matrix and Rotman lenses, which create multiple fixed beams; lens/transmitarray and reflectarray antennas, which focus and steer beams using multiple feeds. These methods are utilized in modern mobile communication base stations due to their capabilities in handling multi-path and wide beam coverage [4]–[6].

\begin{figure}[!t]
\centerline{\includegraphics[width=\columnwidth]{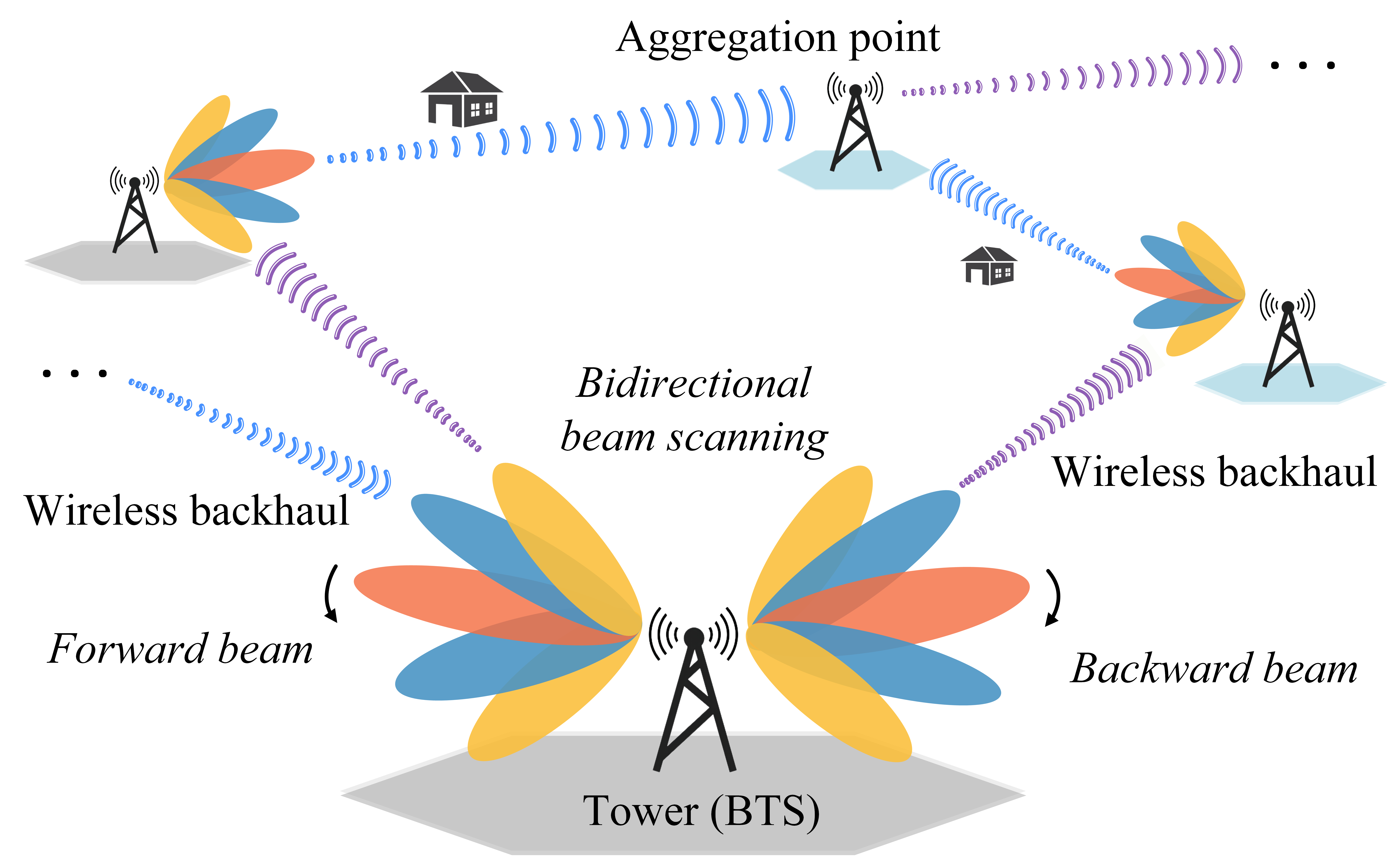}}
\caption{Application scenario of a multibeam base station with bidirectional scanning characteristics in wireless backhaul.}
\label{fig1}
\end{figure}

\begin{figure*}[!t]
\centerline{\includegraphics[width=0.9\textwidth]{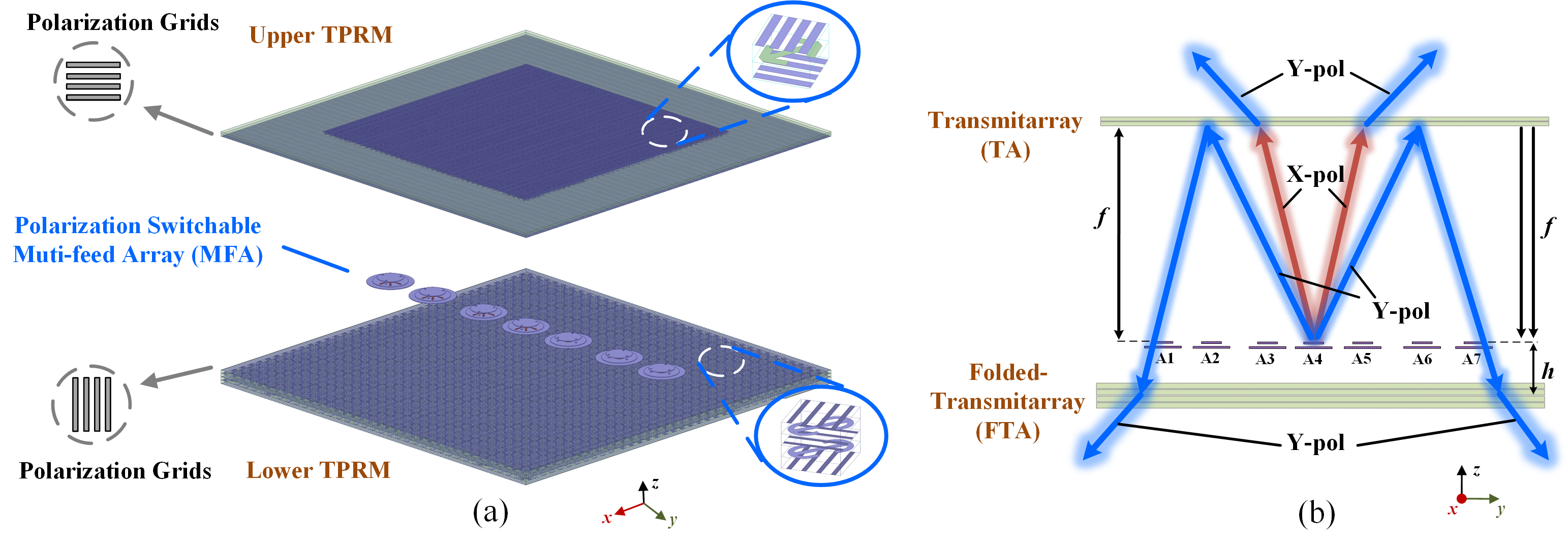}}
\caption{(a) Isometric view of the bidirectional multibeam antenna. (b) Schematic diagram of bidirectional beam scanning antenna and radiation in various working states.}
\label{fig2}
\end{figure*}

A technique to address the hardware complexity of large phased arrays is to use a dielectric lens [7]–[9]. There are various methods for achieving beam scanning with a dielectric lens antenna, including zoned lens [11], [12], Luneburg lens [13]–[15], multi-focal lens [16], [17]. Additionally, a lens in the form of a transmitarray can achieve robust beam-scanning capability. [18] employs the principle of single-focus phase distribution and a single-feed antenna for 2-D electronic beam-steering. Nevertheless, this design is not suitable for lenses that are illuminated by multiple feeds, resulting in power leakage and ultimately leading to a reduction in gain, particularly under large scanning angles. In this context, the implementation of the bifocal phase design method can significantly enhance the current situation [19]–[21]. In [21], a +/-40° scan range is achieved using dielectric bifocal transmitarray with a scan loss of 1.2 dB. Furthermore, a 2-D electronic beam-scanning transmitarray employing dual-layer reconfigurable Huygens element was reported with a simple biasing network [22]. To increase the beam scanning range, a cylindrical elliptical-shaped transmitarray for +/-45° beam scanning was developed by moving the feed along a focal arc [23].

Whilst numerous lens/transmitarray antennas have been proposed to achieve unidirectional beam scanning, [24] introduced a method for simultaneous bidirectional beam scanning to efficiently transmit and receive signals in two opposite directions. However, this method could not facilitate the independent operation of forward or backward modes. More recently, the transmitarray (TA) and reflectarray (RA) have been combined to achieve bidirectional beam steering. The core of the design strategy is to entail the adjustment of the phase of each radiating element to control the direction of the radiation wave [26], [27]. Transmit-reflect array (TRA) structures are versatile and capable of facilitating bidirectional radiation [28]–[30]. A transmitarray was proposed using tunable resonant layer PIN diodes to achieve bidirectional beam scanning [28]. The challenge associated with the design is incorporating a substantial number of PIN diodes in the antenna aperture, which has rather complex structures, resulting in significant losses. The authors in [29] utilize a phase-shifting metasurface backed by frequency selective surface (FSS) to achieve broadband bidirectional radiation. However, the design requires manual rotation of the feed antenna to switch the radiation state, which greatly reduces the convenience and efficiency of beam switching. Recently, metasurface units with polarization-conversion characteristics can be employed to implement the transmitarray for both linear polarization conversion [31], [33], and circular polarization conversion [32]. By designing the unit cell with a transmission polarization-rotating metasurface and polarizing grids [33], high-gain beams for both forward and backward radiation were achieved.

\begin{figure*}[!t]
\centerline{\includegraphics[width=0.9\textwidth]{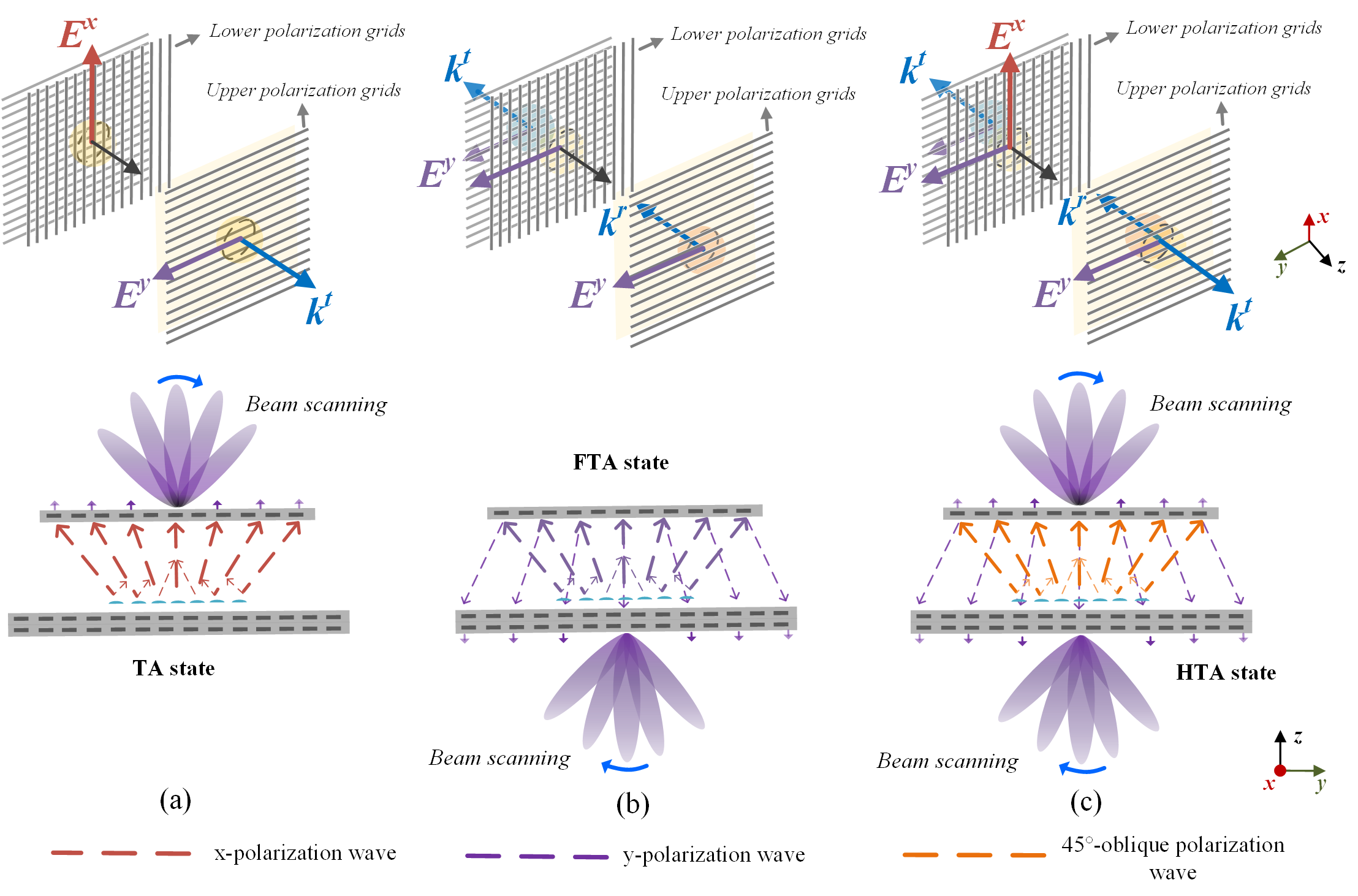}}
\caption{Working principle of reconfigurable bidirectional beams based on polarization switching antenna. (a) Forward beam. (b) Backward beam. (c) Bidirectional beam.}
\label{fig3}
\end{figure*}

Modern wireless communication systems require efficient and adaptable beam scanning capabilities. Bidirectional beam scanning, where the radiation direction can be dynamically adjusted, is becoming increasingly important for applications demanding coverage in both directions. To ensure robust and energy-efficient performance, these systems need antennas that combine high gain for strong signal transmission and reception with a simple structure to minimize power consumption. Fig. 1 demonstrates the application scenario of the bidirectional beam scanning in the base station of the wireless backhaul networks [25]. By offering an extensive beam coverage area, bidirectional beam scanning antennas increase flexibility in network deployment and facilitate the creation of more efficient communication links within the backhaul network. This paper presents a bidirectional multibeam hybrid transmitarray (HTA) based on a polarization rotating metasurface combined with a polarization switchable feed array. The resulting HTA is a combination of traditional TA and folded-transmitarray (FTA). To the best of the authors’ knowledge, the proposed antenna is the first demonstration capable of achieving multi-beam scanning in both unidirectional and bidirectional modes by controlling the polarization states of the feed antenna array.

The rest of the paper is organized as follows. First, the antenna configuration and operating principle of the proposed design are elaborated. Then, a detailed discussion of the design principles of the unit cells and phase compensation, along with the polarization-switchable multi-feed array (MFA), is elucidated. In Section IV, the experimental results are compared, with a discussion of the comparison with existing studies. Finally, conclusions are drawn in Section V.

\section{Working Principle of The Multibeam Hybrid Transmitarray With Bidirectional Radiation}
Fig. 2(a) depicts the proposed bidirectional multibeam HTA, comprising three parts: the upper transmission polarization-rotating metasurface (TPRM), the lower TPRM, and the MFA mounted between the two TPRMs. The proposed TPRM has a polarization-conversion unit in the middle layer sandwiched by two metallic orthogonal polarization grids printed on the dielectric substrate, which can work as a lens or reflector for the incident wave depending on the grid orientation and incident wave polarization. It is noted that the proposed TPRM can rotate the polarization of the transmitted wave with respect to that of the incident waves by 90°. The TPRM serves three essential functions: Firstly, it transmits or reflects the incident wave based on its associated polarization; secondly, it focuses the transmitted wave to generate pencil beams by appropriately adjusting the structure of TPRM unit cells; and thirdly, it imparts a 90-degree polarization rotation to the transmitted wave. The seven feed antennas, labeled A1 to A7, within the MFA, are symmetrically positioned along the y-axis between the upper and lower TPRMs.

Fig. 3 illustrates the operating principle of multi-mode radiation (unidirectional and bidirectional radiation), and the incident waves generated by MFA in the three states are x-polarization, y-polarization, and 45°-oblique polarization, respectively. The upper and lower polarization grids closest to the MFA are oriented parallel to the y-axis and x-axis, respectively. It is worth noting that the lower metasurface is composed of two TPRMs, which can facilitate double polarization rotation to obtain the same polarization for the forward, backward, and bidirectional transmitted beams.

When illuminated by an x-polarized wave from the feed antenna, the upper TPRM will transmit, focus, and rotate the x-polarized incident waves to a y-polarized pencil beam with high directivity in the forward direction (+z direction, TA state), as shown in Fig. 3(a). The y-polarized waves emitted by the MFA undergo complete reflection at the upper TPRM's metallic grid. The lower TPRM then transmits and focuses these reflected waves, generating backward radiation along the -z direction (FTA state), as shown in Fig. 3(b). To achieve simultaneous bidirectional radiation, the polarization of feed antennas is switched to either +45° or 45° relative to the x-axis, as illustrated in Fig. 3(c). In this state, the incident wave generated by the feed antenna can be decomposed into two orthogonal polarization components (x- and y-polarization) with the same magnitude. The upper TPRM transmits and focuses the x-polarized component while the y-polarized component undergoes the same process as the FTA state. This operation can be regarded as the combination of the TA and FTA along two opposite directions, thus denoted as the HTA state. By altering the polarization of the MFA using a reconfigurable feed network, forward, backward, and bidirectional beam scanning performance can be achieved.

\section{Design of The Proposed Bidirectional Rotating Mutibeam Hybrid Transmitarray}
\subsection{Unit Cell Design}
The antenna element for the upper TPRM is donated as Unit Cell 1 (UC1), as shown in Fig. 4(a). The central metallic layer features a double arrow-shaped strip designed to convert the polarization state. Two orthogonal metallic wire-grid polarization grids are positioned on the top and bottom layers, with the wire orientations aligned parallel to the x-axis and y-axis, respectively [20]. The polarization grids exhibit polarization selectivity, allowing only the component of an incident electromagnetic wave with a perpendicular polarization to pass through while reflecting the component with a parallel polarization direction. Therefore, when the electromagnetic wave is illuminated from bottom to top on UC1, it can transform the x-polarization wave into a y-polarization wave while the y-polarization wave is reflected. The dielectric substrate of UC1 is F4B with ${\varepsilon _{\rm{r}}}$ of 2.65 and a loss tangent of 0.001. Each dielectric layer has a thickness of ${{\rm{h}}_{1}}$. As shown in Fig. 4, the polarization conversion layer consists of a rectangular metal strip with an inclined angle of 45° with respect to the x-axis. The period of the UC1 is set as ${{\rm{P}}_{1}}$ = 6 mm, corresponding to 0.2$\lambda$. To achieve a high Polarization Conversion Rate (PCR) and efficiently convert polarization across the desired frequency band, the design parameters were optimized using commercial software HFSS. Master-slave boundary conditions and Floquet ports were employed within the simulation process. The total thickness of UC1 is 4 mm, corresponding to 0.13$\lambda$, where $\lambda$ is the free space wavelength at the center frequency of 10 GHz.

\begin{figure}[!t]
\centerline{\includegraphics[width=\columnwidth]{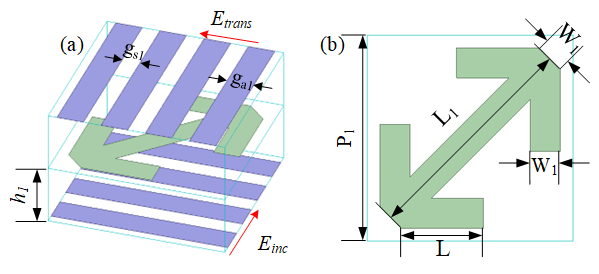}}
\caption{Transmitarray unit cell UC1. (a) 3D view. (b) Metallic pattern in the middle layer. The geometrical parameters are: ${{\rm{g}}_{a1}}$ = 0.9 mm, ${{\rm{g}}_{s1}}$ = 0.6 mm, ${{\rm{L}}_{1}}$ = 6.56 mm, ${{\rm{W}}_{1}}$ = 0.88 mm, and ${{\rm{h}}_{1}}$ = 2 mm.}
\label{fig4}
\end{figure}

\begin{figure}[!t]
\centerline{\includegraphics[width=\columnwidth]{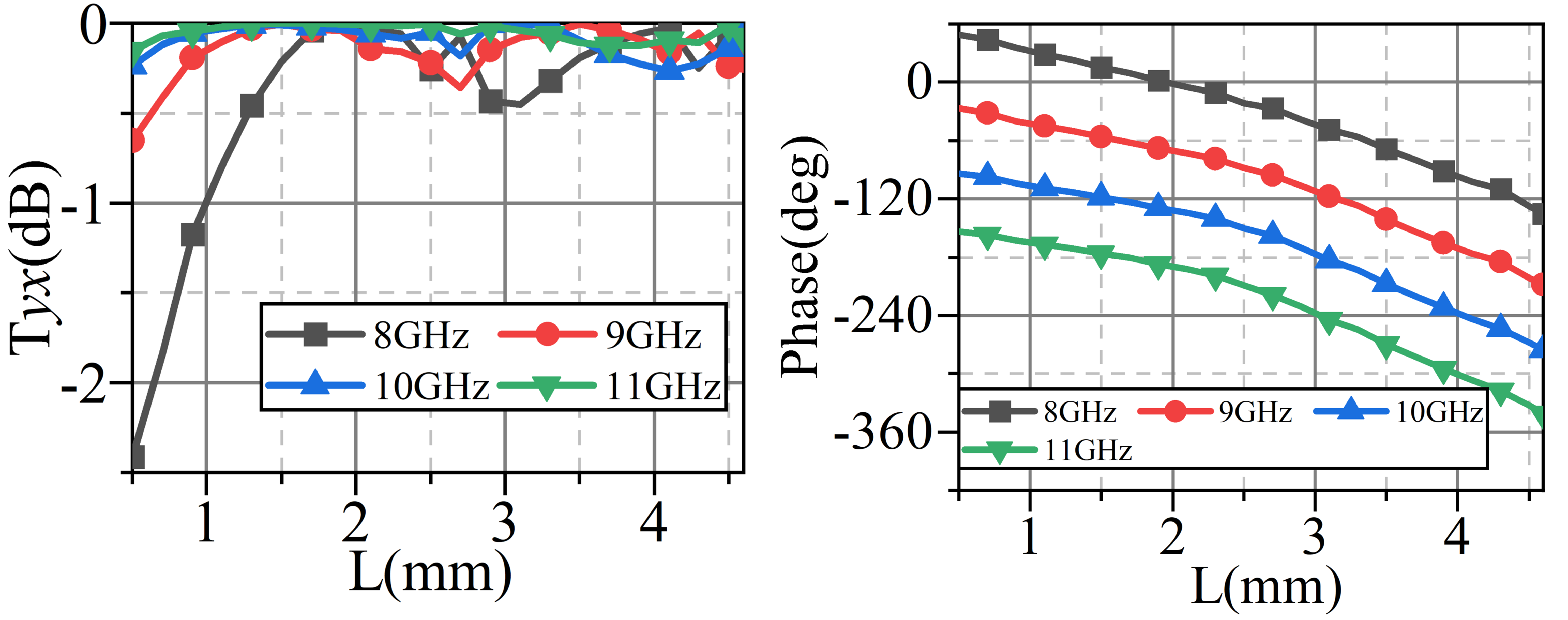}}
\caption{Transmission magnitude and phase versus the parameter of “L” at different frequencies.}
\label{fig5}
\end{figure}

\begin{figure}[!t]
\centerline{\includegraphics[width=\columnwidth]{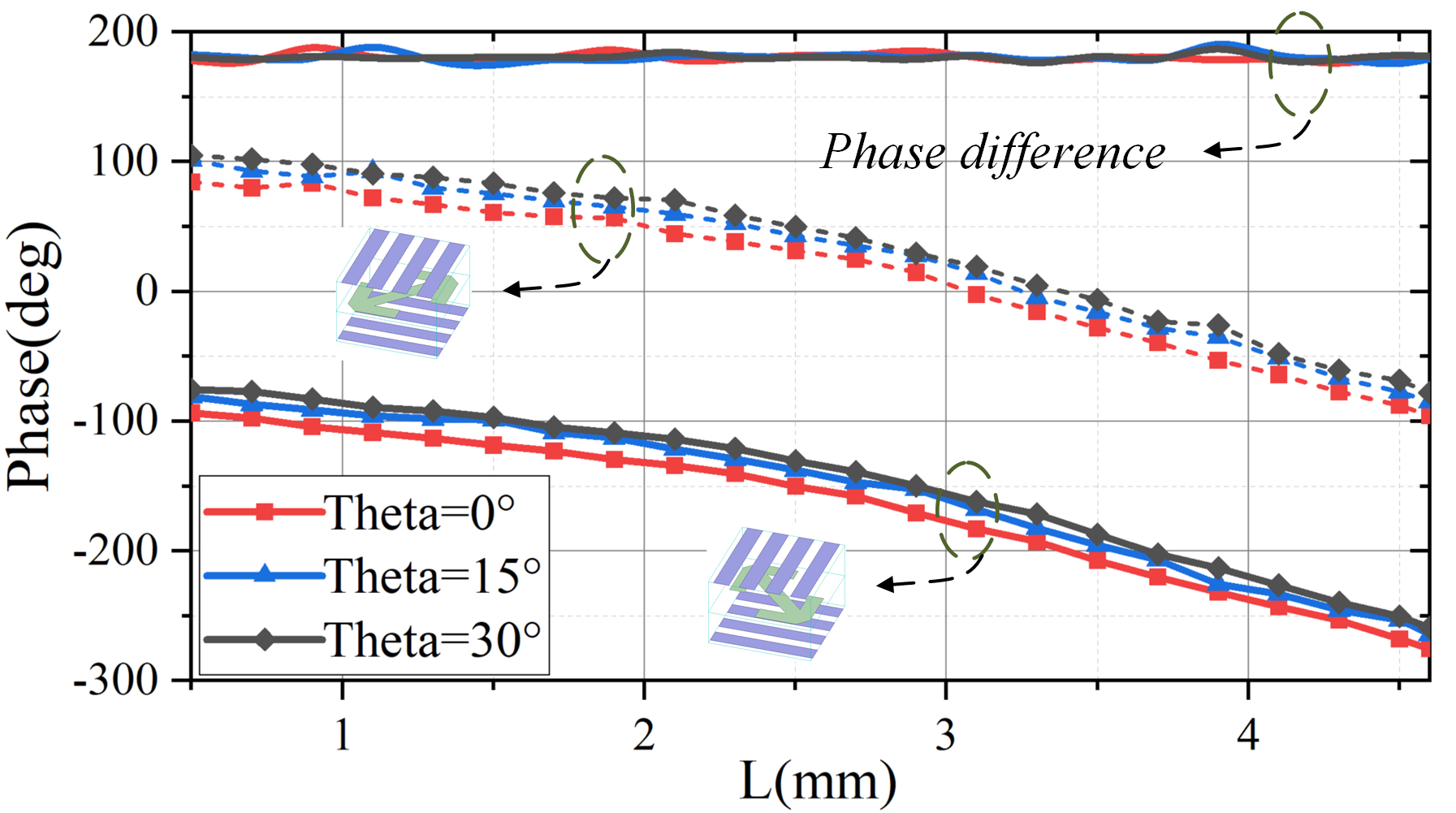}}
\caption{Phase versus “L” of UC1 at 10 GHz at vertical and oblique incidence.}
\label{fig6}
\end{figure}

\begin{figure}[!t]
\centerline{\includegraphics[width=\columnwidth]{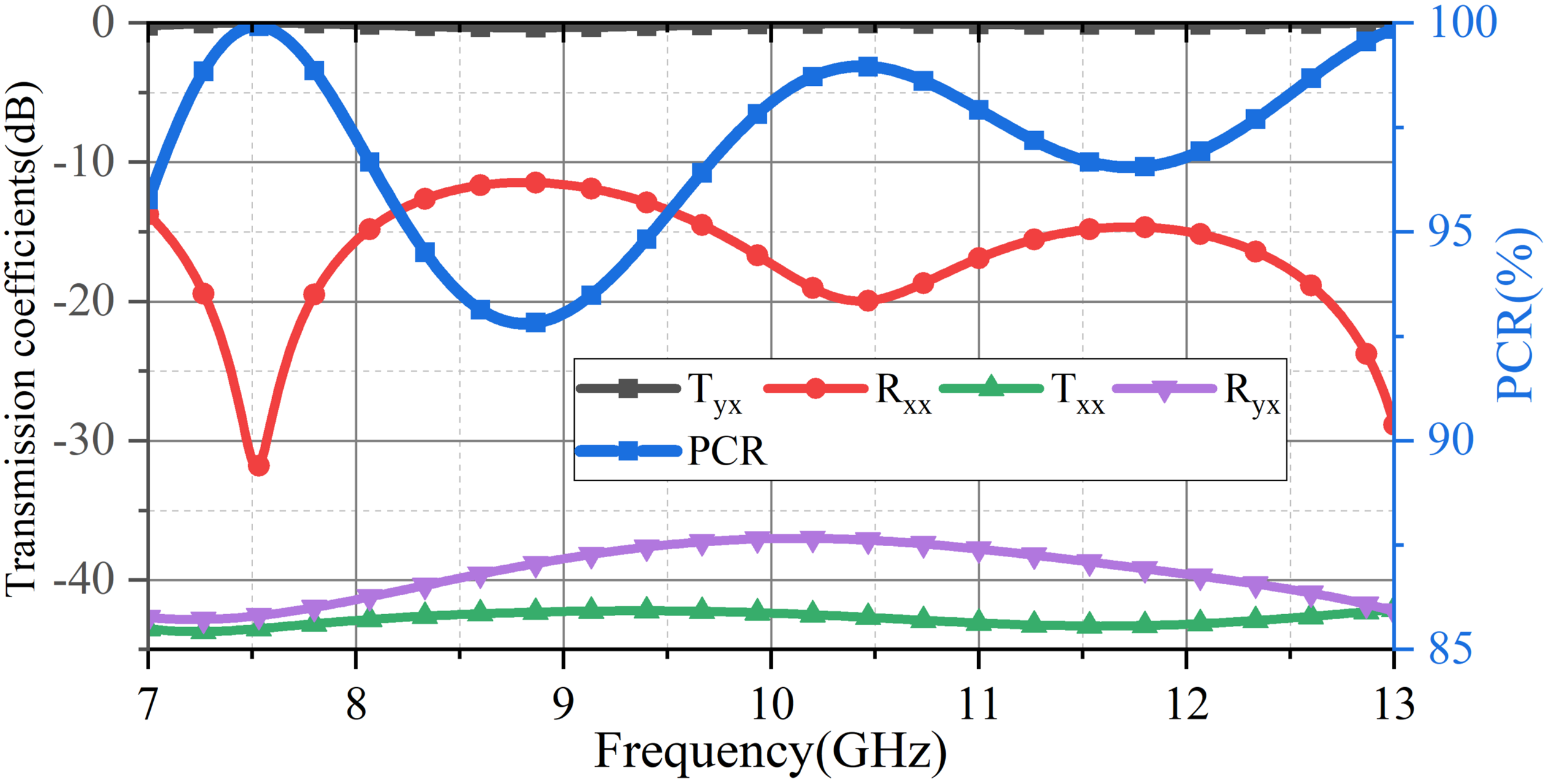}}
\caption{Transmission, reflection coefficients and PCR curves of UC1 when L = 2.4 mm.}
\label{fig7}
\end{figure}

To compensate for the phase delay resulting from diverse feeding positions and beam directions, it is imperative to employ unit cells featuring a 360° phase tuning range and high transmission efficiency. Fig. 5 shows the relationship between the simulated magnitude and phase of the transmission coefficient of UC1 with the variation of the parameter “L” at different frequencies. As the parameter “L” varies from 0.5 mm to 4.6 mm, a phase variation of 180° is attained, and the $\left| {{T_{yx}}} \right|$ is close to 0 dB. The phase shift curves are almost parallel for frequencies from 8 to 11 GHz, showing wideband properties. Fig. 6 illustrates the phase curve for both vertical and oblique incidences at 10 GHz. The overall phase deviation for oblique incidence is minimal, and rotating the polarization conversion layer by 90° introduces an additional 180° phase shift. Consequently, the phase variation of UC1 can cover the entire 360° phase range. The PCR of the proposed unit is defined as
\begin{equation}PCR = |T_{yx}|^2/(|T_{yx}|^2 + |T_{xx}|^2 + |R_{yx}|^2 + |R_{xx}|^2)\label{eq}\end{equation}\noindent where ${{T_{yx}}}$ and ${{T_{xx}}}$ are defined as the transmission coefficients, and ${{R_{yx}}}$ and ${{R_{xx}}}$ are defined as the reflection coefficients, the first in the subscript denotes the polarization of the transmitted or reflected wave while the second indicates the polarization of the incident wave. Due to the strong polarization selection characteristic of UC1, it can be seen that ${{T_{yx}}}$ is close to 0 dB in the range of frequency bands. As depicted in Fig. 7, the PCR of UC1 exceeds 92.8$\%$ across the frequency range from 7 GHz to 13 GHz.

To ensure the same polarization state (y-polarization) for both forward and backward beams, a double polarization-conversion unit is stacked to form the FTA Unit Cell 2 (UC2), as illustrated in Fig. 8. UC2 can be considered a combination of two stacked UC1 with adjusted dimensions to achieve the same polarization as in the forward radiation mode. The unit period of UC2 is optimized as ${{\rm{P}}_{2}}$ = 10 mm. The dielectric material of UC2 is the same as that of UC1 (${\varepsilon _{\rm{r}}}$ = 2.65). The total thickness of UC2 is 12 mm, approximately equivalent to 0.4$\lambda$ wavelengths.

\begin{figure}[!t]
\centerline{\includegraphics[width=\columnwidth]{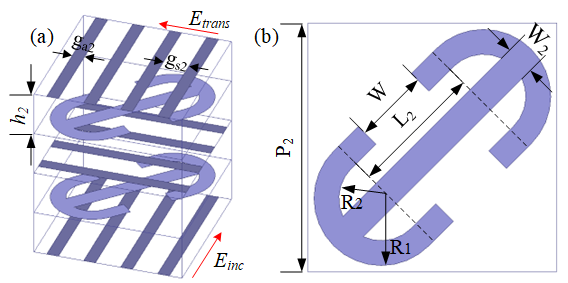}}
\caption{Folded-transmitarray unit cell UC2. (a) 3D view. (b) Metallic pattern in the middle layer. The geometrical parameters are: ${{\rm{g}}_{a2}}$ = 0.8 mm, ${{\rm{g}}_{s2}}$ = 1.7 mm, ${{\rm{R}}_{1}}$ = 2.7 mm,  ${{\rm{R}}_{2}}$ = 1.7 mm,  ${{\rm{W}}_{2}}$ = 1.15 mm, ${{\rm{L}}_{2}}$ = 5.8 mm, and ${h_2}$ = 4 mm.}
\label{fig8}
\end{figure}

\begin{figure}[!t]
\centerline{\includegraphics[width=\columnwidth]{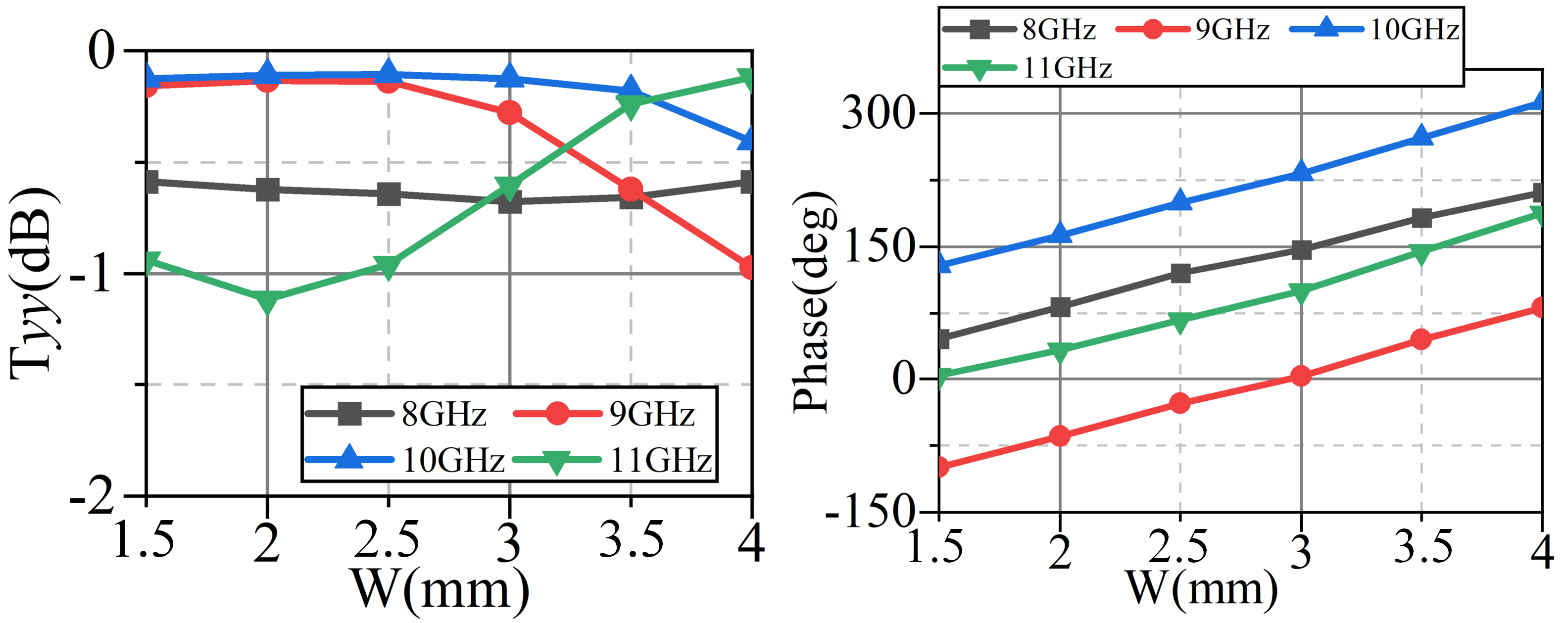}}
\caption{Transmission magnitude and phase versus “W” at different frequencies.}
\label{fig9}
\end{figure}

\begin{figure}[!t]
\centerline{\includegraphics[width=\columnwidth]{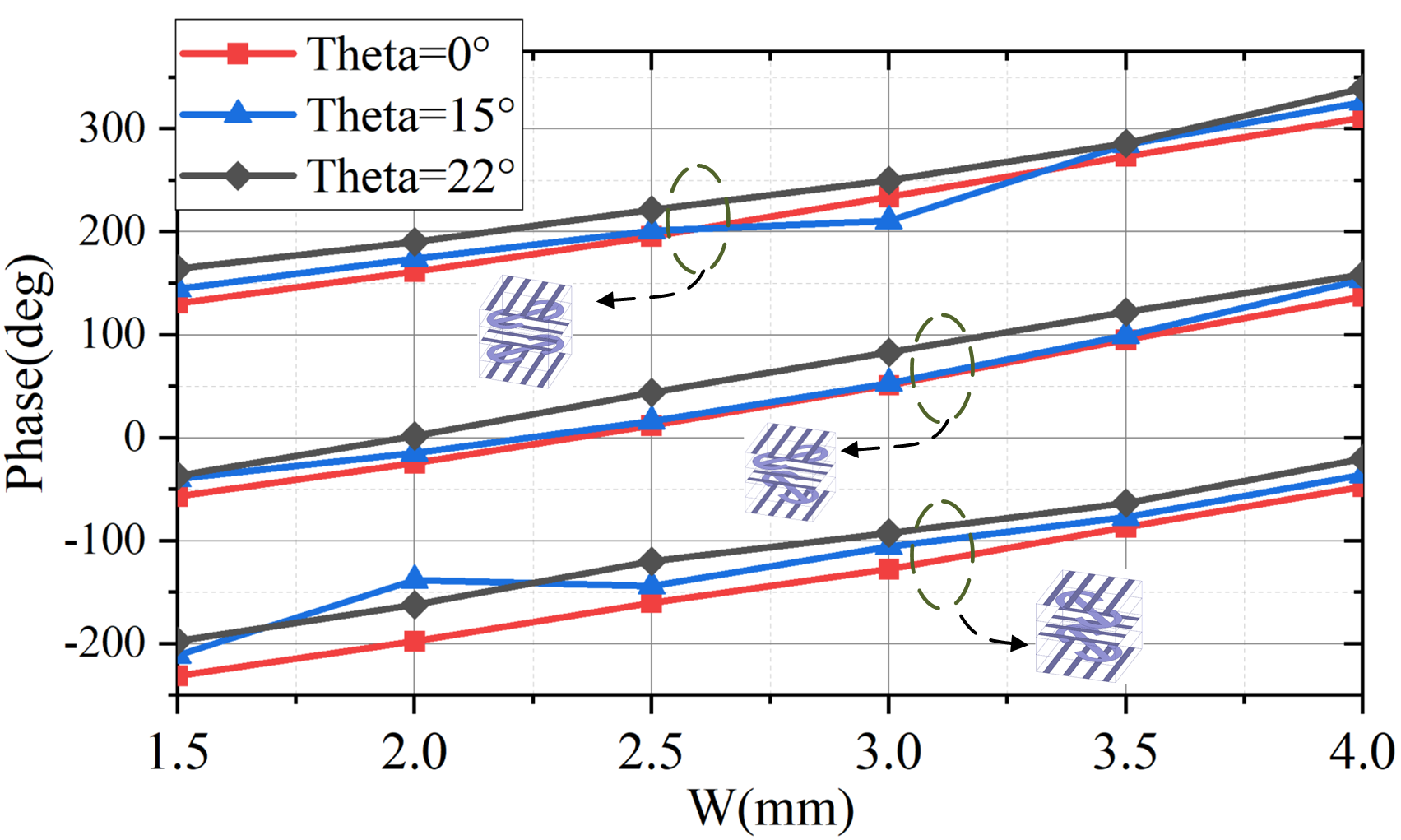}}
\caption{Phase versus “W” of UC2 at 10 GHz at vertical and oblique incidence.}
\label{fig10}
\end{figure}

\begin{figure}[!t]
\centerline{\includegraphics[width=\columnwidth]{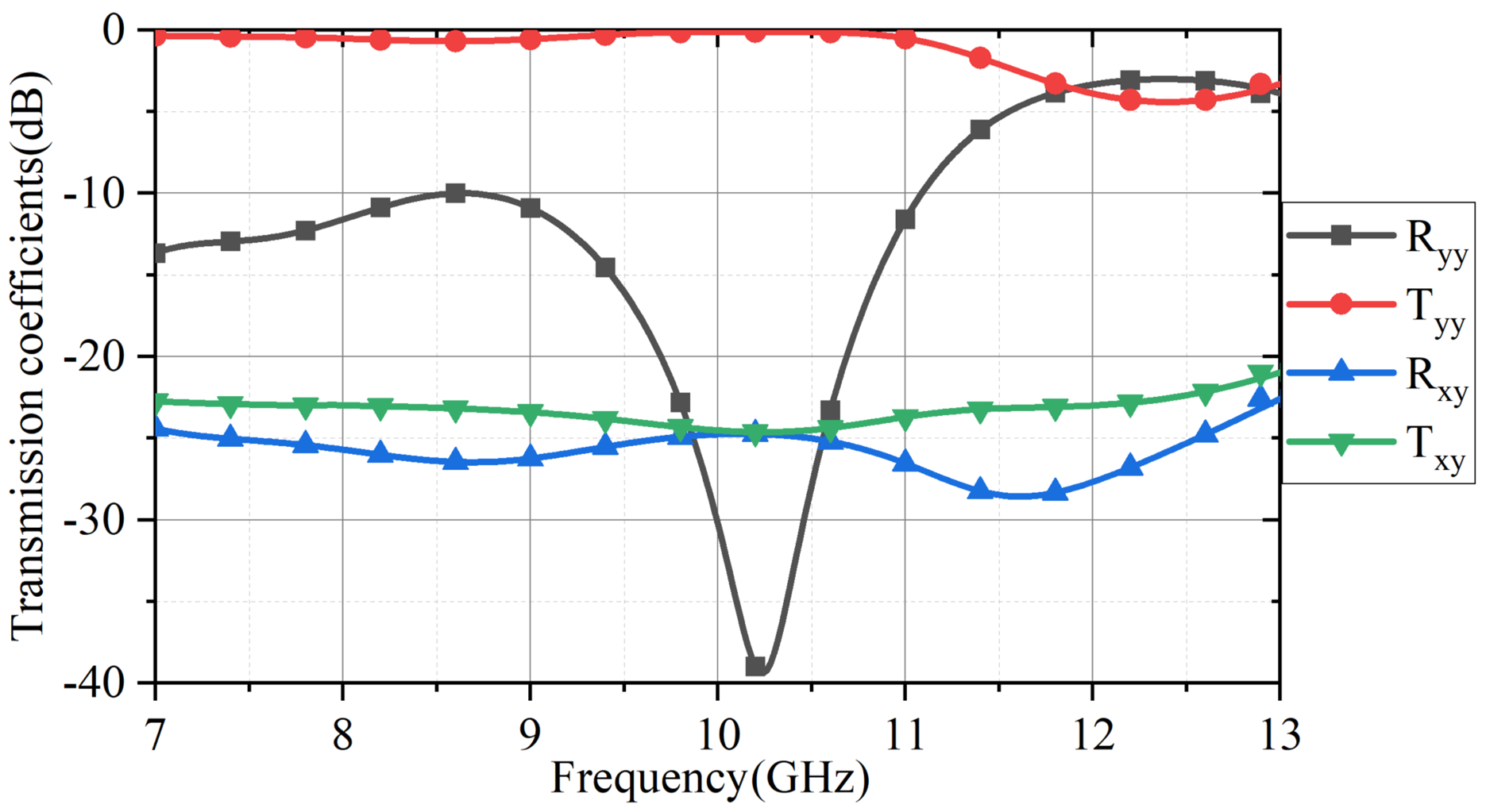}}
\caption{Transmission and reflection coefficients of UC2 when W = 3 mm.}
\label{fig11}
\end{figure}

\begin{figure*}[!t]
\centerline{\includegraphics[width=0.9\textwidth]{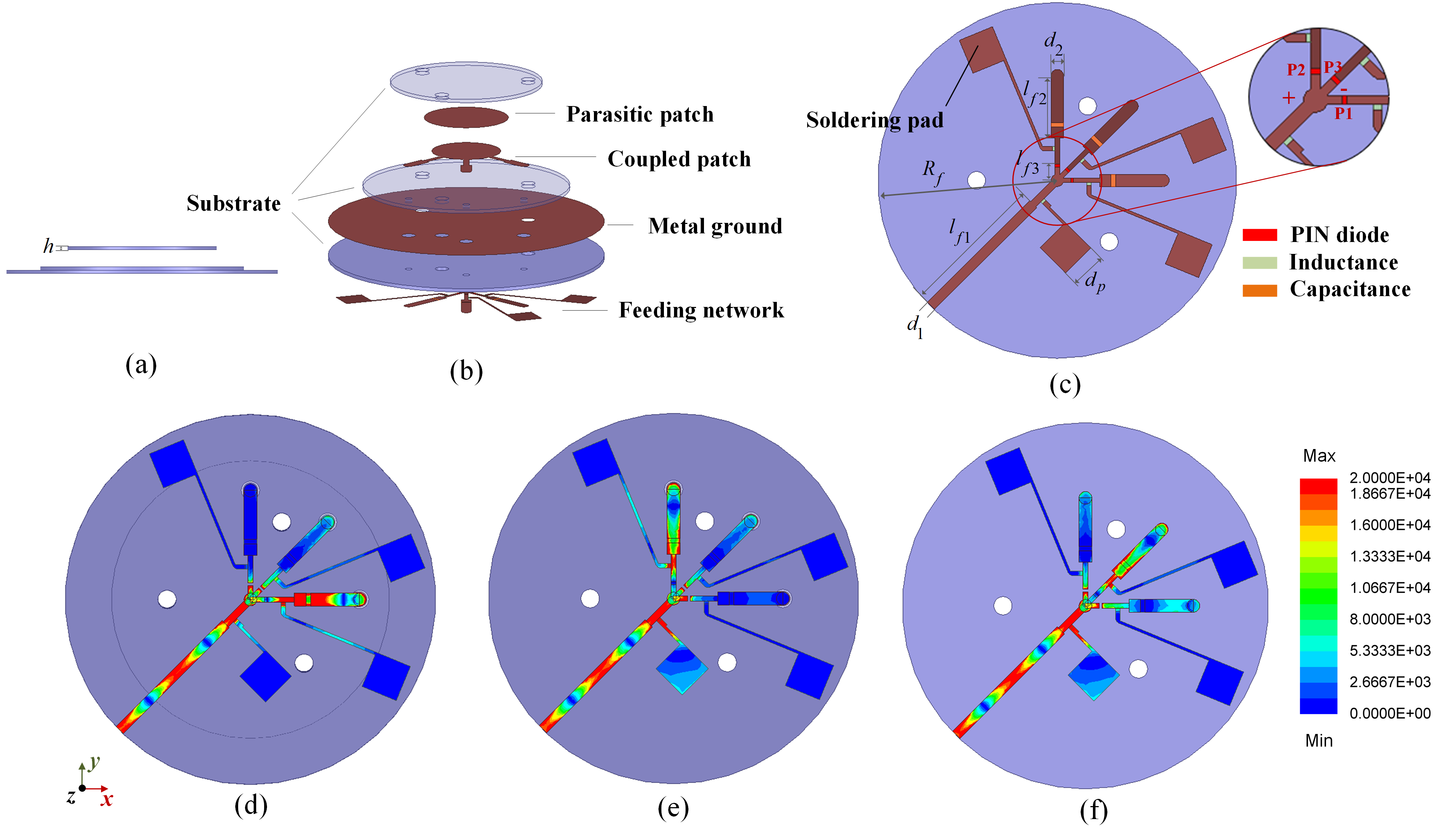}}
\caption{(a) Side view of feed antenna. (b) Exploded view. (c) The feed network of the feed antenna. (d) The electric field distribution in x-polarization. (e) The electric field distribution in y-polarization. (f) The electric field distribution in 45°oblique-polarization. The geometrical parameters are: ${{\rm{R}}_{f}}$ = 20 mm, ${{\rm{d}}_{p}}$ = 4 mm, ${{\rm{l}}_{f1}}$ = 16.15 mm, ${{\rm{l}}_{f2}}$ = 7.04 mm, ${{\rm{d}}_{1}}$ = 1.09 mm, ${{\rm{d}}_{2}}$ = 1.39 mm, ${{\rm{l}}_{f3}}$ = 1.5 mm, ${h}$ = 2.5 mm.}
\label{fig12}
\end{figure*}

\begin{figure}[!t]
\centerline{\includegraphics[width=\columnwidth]{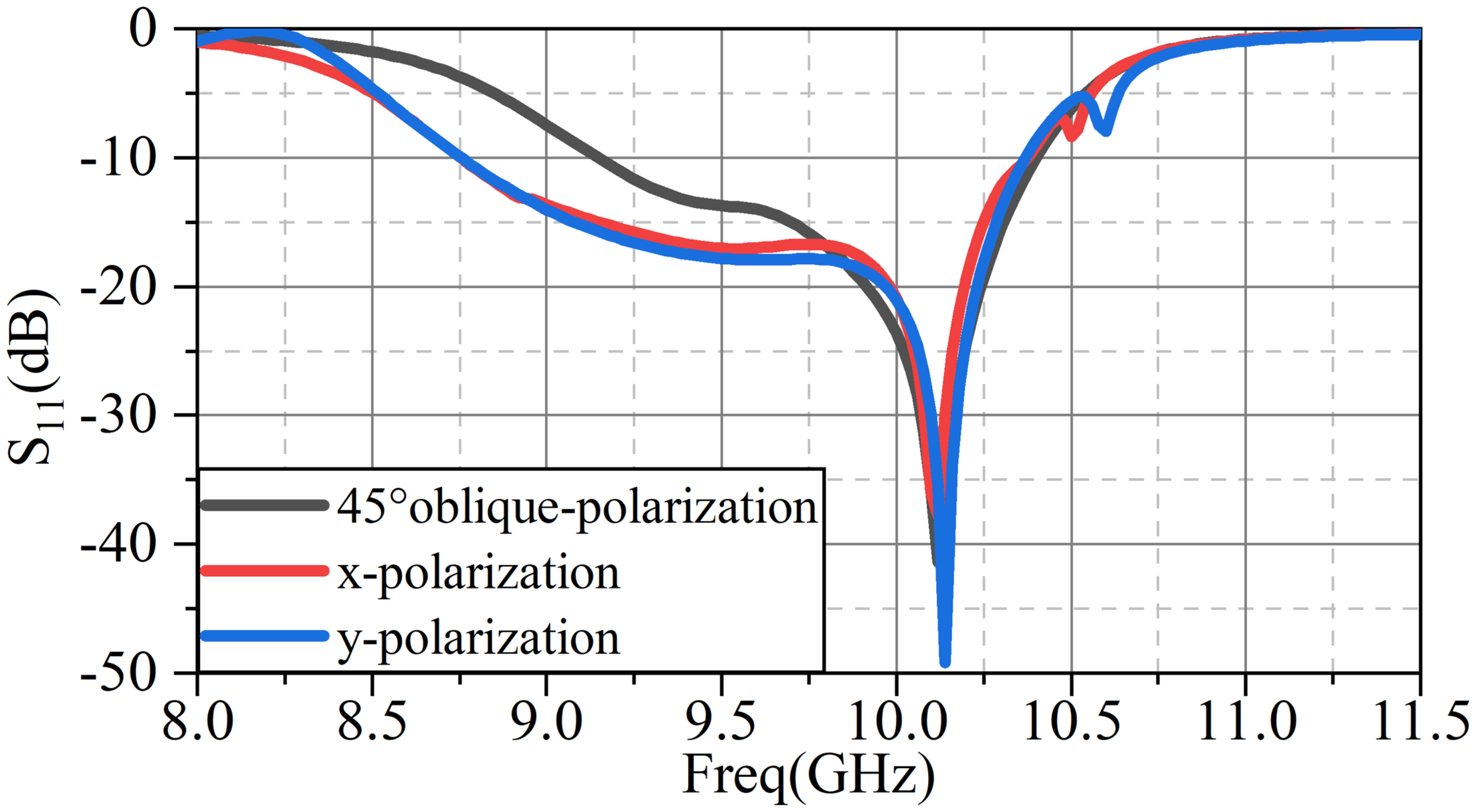}}
\caption{The simulated ${{\rm{S}}_{11}}$ parameters of feed antenna in different polarization states.}
\label{fig13}
\end{figure}

\begin{figure}[!t]
\centerline{\includegraphics[width=\columnwidth]{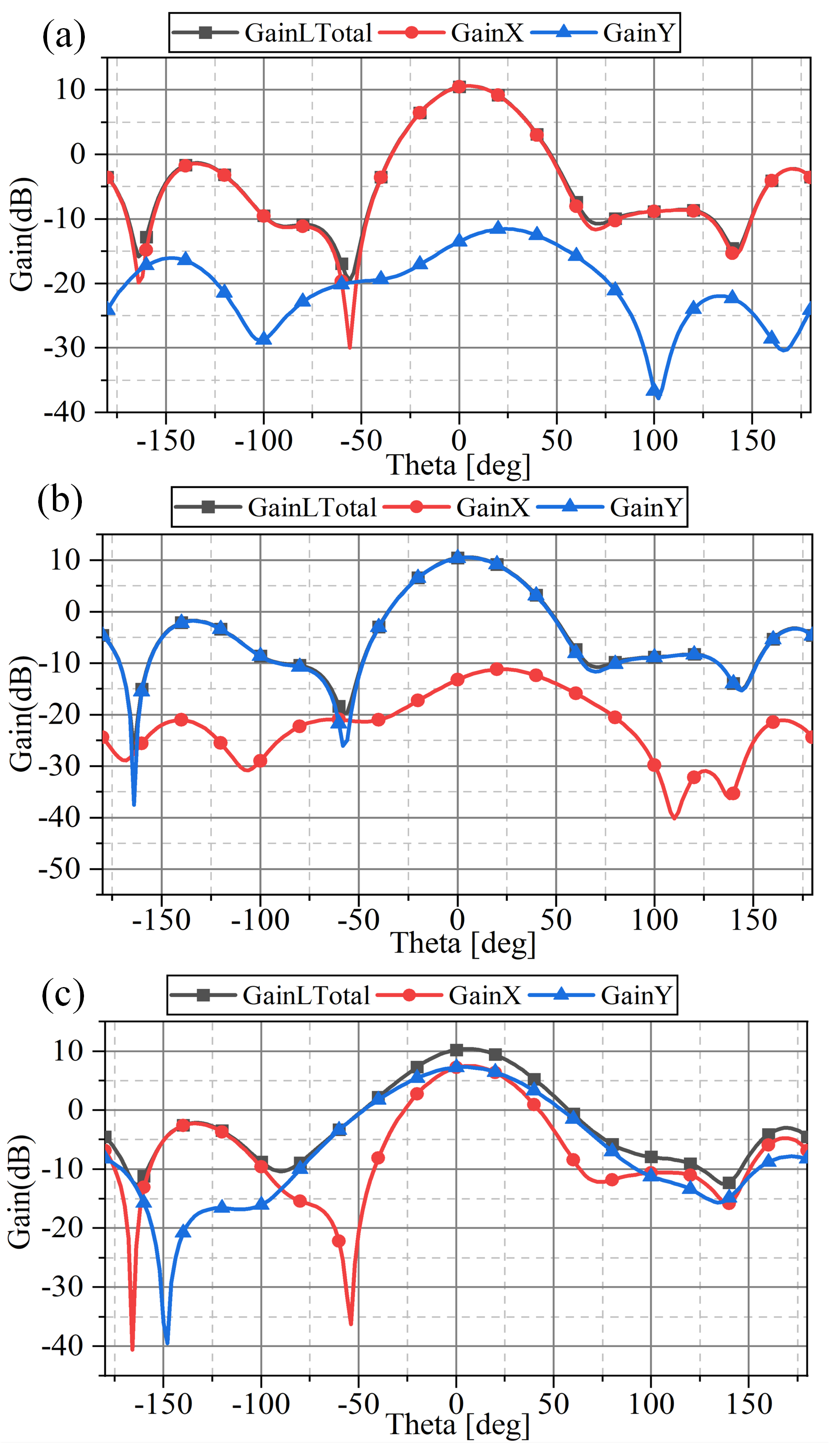}}
\caption{Radiation patterns of the feed antenna. (a) x-polarization. (b) y-polarization. (c) 45°oblique-polarization.}
\label{fig14}
\end{figure}

\begin{figure}[!t]
\centerline{\includegraphics[width=\columnwidth]{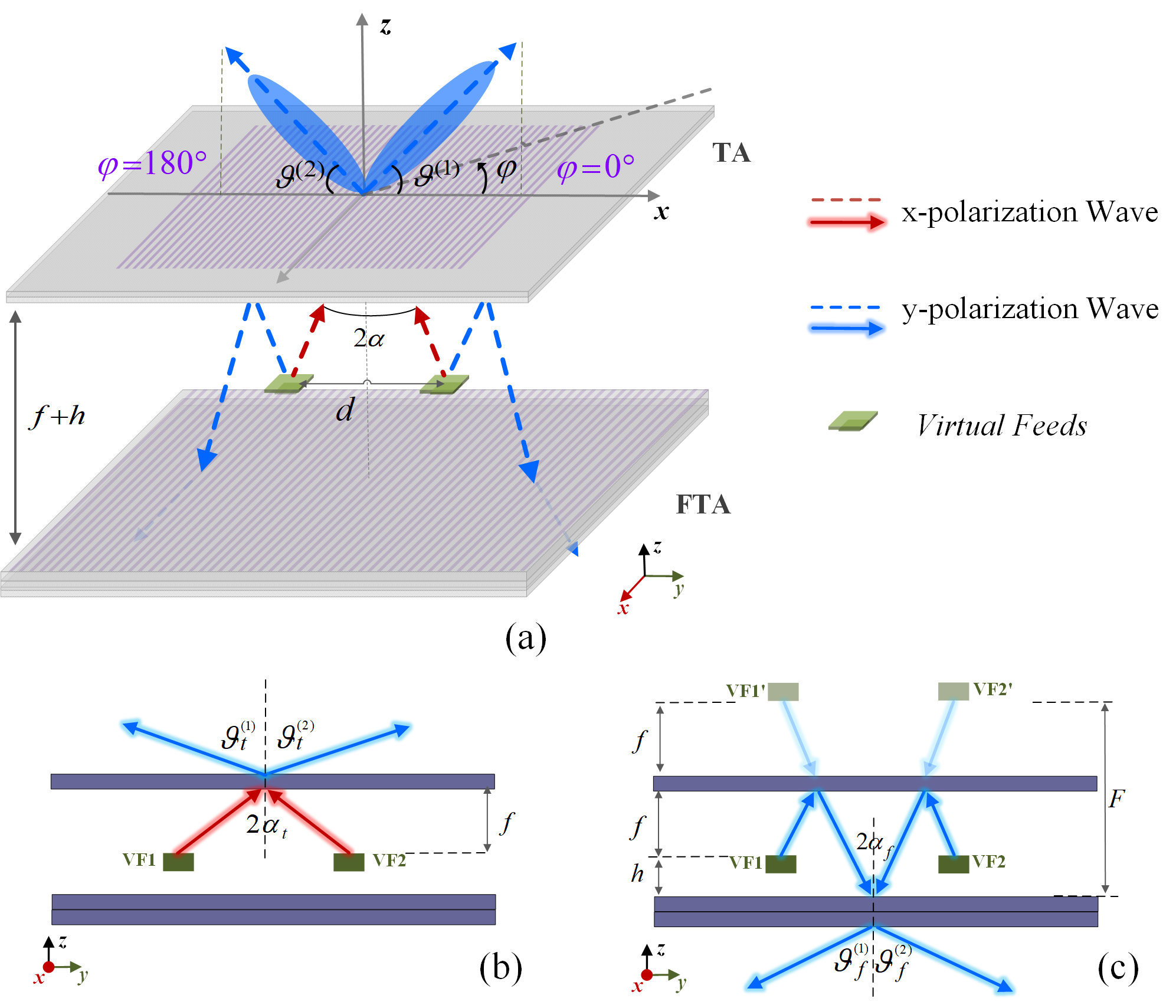}}
\caption{Schematic diagram of bifocal phase design principle. (a) Isometric view. (b) Forward radiation. (c) Backward radiation.}
\label{fig15}
\end{figure}

\begin{figure}[!t]
\centerline{\includegraphics[width=\columnwidth]{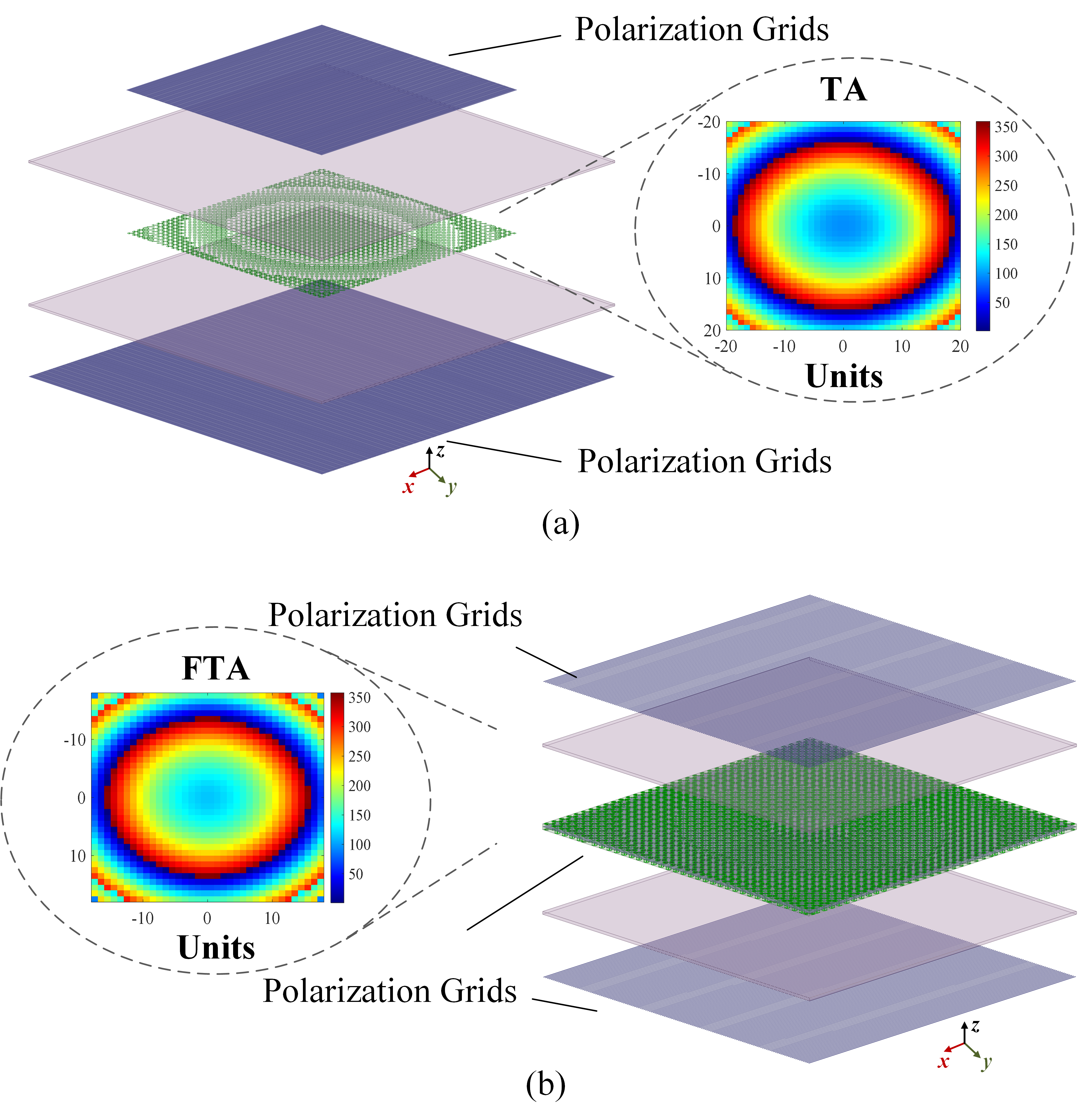}}
\caption{Geometry of the model and calculated phase distributions on (a) TA, and (b) FTA.}
\label{fig16}
\end{figure}

\begin{figure}[!t]
\centerline{\includegraphics[width=\columnwidth]{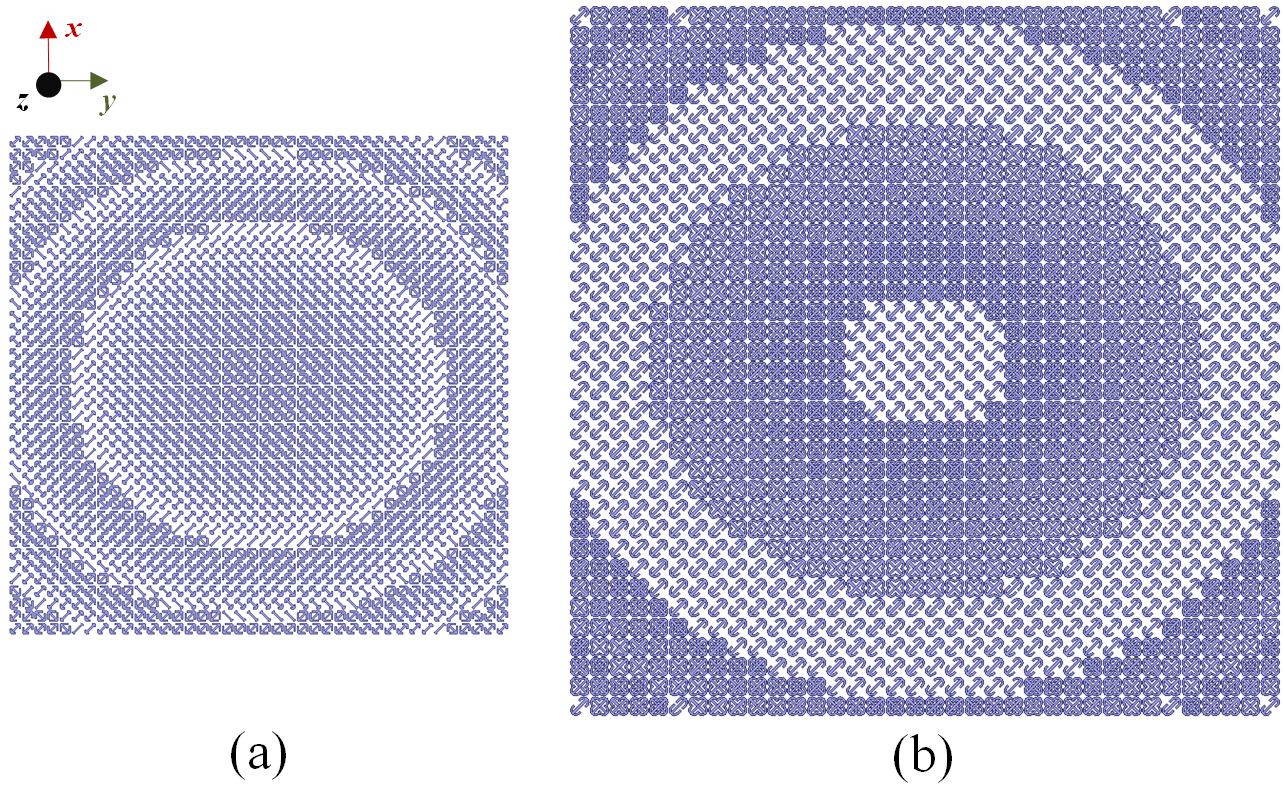}}
\caption{Metasurface metal pattern layer structure. (a) Metasurface of upper TPRM. (b) Metasurface of lower TPRM.}
\label{fig17}
\end{figure}

Fig. 9 shows the simulated magnitude and phase of the transmission coefficient of UC2 at different frequencies. As the parameter “W” varies from 1.5 mm to 4 mm, a phase variation of 180° is achieved, and the $\left| {{T_{yy}}} \right|$ is higher than -1.1 dB. As shown in Fig. 10, UC2 exhibits nearly parallel phase shift curves across various frequencies, leading to broad bandwidth and efficient phase compensation. Additionally, the low slope of the phase curve minimizes sensitivity to manufacturing tolerances. This reduces unwanted phase shifts caused by potential errors from manufacturing UC2 and maintains stable performance even at oblique angles. The element phase remains relatively constant with increasing incidence angles. Similar to the approach used in UC1, rotating either the top or bottom half of UC2 by 90 degrees introduces a 180-degree phase shift compared to the original phase response. This manipulation allows UC2 to achieve phase modulation across the entire 360-degree range. Fig. 11 depicts the transmission and reflection coefficients of UC2.

\subsection{Design of MFA}
The multi-mode HTA relies on polarization-switchable microstrip antenna elements. Fig. 12(a) depicts the structure of the designed polarization switchable antenna, which consists of three 0.5 mm-thick substrates of RO4003, a metal ground, a coupled patch with ${{\rm{R}}_{3}}$ = 5 mm, a parasitic patch with ${{\rm{R}}_{4}}$ = 6.1 mm, and a polarization switching feed network. The presence of a parasitic patch positioned above the antenna induces an additional resonance. The dimensions of the parasitic patch are optimized to enhance the bandwidth of the feed antenna.

The three feed lines are connected to the main feed line, as shown in Fig. 12(c). Three PIN diodes (MADP-00090714020x) P1-P3 are employed, and the inductance in the feeding network is 3.3 nH (LQW15AN3N3G8ZD), while a capacitor with a capacity of 20 nF (GRM21B5C1H223JA01L) is used [22]. DC bias lines are soldered to designated pads on the feed network branches, enabling individual control of the polarization states radiated by the antenna. Fig. 12(d)-(f) shows the electric field distribution of the feed antenna at different polarization states.
 
Fig. 13 presents the ${{\rm{S}}_{11}}$ parameters of the feed antenna in different working states. The ${{\rm{S}}_{11}}$ remains below -10 dB between 9.2 GHz and 10.4 GHz. Fig. 14 indicates that the gains of the feed antenna in x-polarization, y-polarization, and 45° oblique-polarization are 10.6 dBi, 10.4 dBi, and 10.5 dBi, respectively. Furthermore, the cross-polarization levels for both the x-polarization and y-polarization states are consistently below -20 dB.

\subsection{Design of Phase Distribution}

It is challenging to optimize the phase distribution for different scanning angles. In single-focus phase designs, the phase mismatch increases with the increase of scan angle, resulting in limited scanning range, gain reduction, and sidelobe degradation. The bifocal principle is introduced to obtain phase compensation for the final HTA to improve the performance of the scanned beams. The operating diagram is shown in Fig. 15; the virtual feed1 (VF1) and virtual feed2 (VF2) are symmetrically placed in the x-z plane at the same height as the MFA, with spacing \textit{f} and \textit{h}, respectively, between TA and FTA.

VF1 and VF2 have an offset angle of $\alpha$ degree along the central axis. When only one of the virtual feeds works, it is the eccentric single-focus metasurface, and its phase compensation distribution can be calculated as follows
\begin{equation}\phi _{ij}^{(f)} = {k_0}(R_{ij}^{(f)} - \sin {\vartheta ^{(f)}}({x_i}\cos {\varphi ^{(f)}} + {y_j}\sin {\varphi ^{(f)}}))\label{eq}\end{equation}\noindent where \textit{f} represents the corresponding feed excitation, $\vartheta$ represents the radiation polar angle of the beam and the metasurface, and $\varphi$ is the rotation angle of the spherical coordinate system. ${R_{ij}}$ is the distance from the virtual feeds to the $(i, j)$-th unit cell.

Unlike the single-focus lens antenna, the phase compensation design of the bifocal metasurface must consider the deflection beam generated by two feeds at different positions. When VF1 works, the polar angle of the scanning beam generated by the metasurface is $(\vartheta = {\vartheta ^{(1)}},\varphi = {0^ \circ })$; similarly, when VF2 works, the polar angle of the beam is $(\vartheta = {\vartheta ^{(2)}},\varphi = {0^ \circ })$. In both cases, the phase compensation distribution across the aperture is
\begin{equation}\begin{array}{l}
\phi _{ij}^{(1)} = {k_0}(R_{ij}^{(1)} - {x_i}\sin {\vartheta ^{(1)}})\\
\phi _{ij}^{(2)} = {k_0}(R_{ij}^{(2)} - {x_i}\sin {\vartheta ^{(2)}})
\end{array}\label{eq}\end{equation}\noindent where ${x_i}$ is the abscissa of the $(i, j)$-th metasurface unit, and ${k_0}$ is the wavenumber in free space.

Since the bifocal antenna system shown in Fig. 15 is placed symmetrically about the z-axis, the two beam deflection angles are equal, that is, $\vartheta = {\vartheta ^{(1)}} = {\vartheta ^{(2)}}$, and the compensation phase required for the $(i, j)$-th unit can be computed as the mean value of two different phase distributions, as shown in the following function:
\begin{equation}{\phi _{bifocal}} = \frac{{\phi _{ij}^{(1)} + \phi _{ij}^{(2)}}}{2} = \frac{{{k_0}(R_{ij}^{(1)} + R_{ij}^{(2)})}}{2}\label{eq}\end{equation}\noindent For the forward radiation (+z), the focal length is the distance from the TA aperture to the virtual feeds, denoted by \textit{f}. For backward radiation (-z), due to the folded structure, the positions of the virtual focal points VF1' and VF2' are mirror images above TA, as shown in the figure below. Therefore, the focal length is expressed by $F = 2f + h$.

The relationship between the focal length, aperture dimension, and the edge taper can be followed by equation (5) [39], where \textit{D} is the lateral size of the antenna. To achieve an optimum aperture efficiency, the \textit{f}/\textit{D} is determined by using the ${\alpha _{ - 10 dB}}$ (-10 dB taper of the feed antenna) to obtain the optimal aperture efficiency.
\begin{equation}f = \frac{D}{{2\tan ({\alpha _{ - 10 dB}})}}\label{eq}\end{equation}\noindent Based on the above design theory, the focal lengths of \textit{f} and \textit{F}  are optimized to 171 mm and 384 mm, respectively. The interval \textit{d} between the two virtual feeds was chosen to be 220 mm. To reflect as many electromagnetic waves produced by the feed antenna as possible, the size of the physical aperture of the upper TPRM is chosen the same as that of the lower TPRM. The phase distribution is illustrated in Fig. 16, and the upper and lower metasurface metal layer is designed as shown in Fig. 17.

To demonstrate the operation of the boresight beam, the electric E-field distributions and radiation patterns for x-polarization, y-polarization, and 45° oblique-polarization, respectively, are shown in Fig. 18. It is obvious that quasi-spherical waves are successfully converted into plane waves by the upper and lower TPRMs. Moreover, it is confirmed from the far-field patterns that the resultant antenna can generate unidirectional beams for the TA and FTA states and a bidirectional beam for the HTA state.

\begin{figure}[!t]
\centerline{\includegraphics[width=\columnwidth]{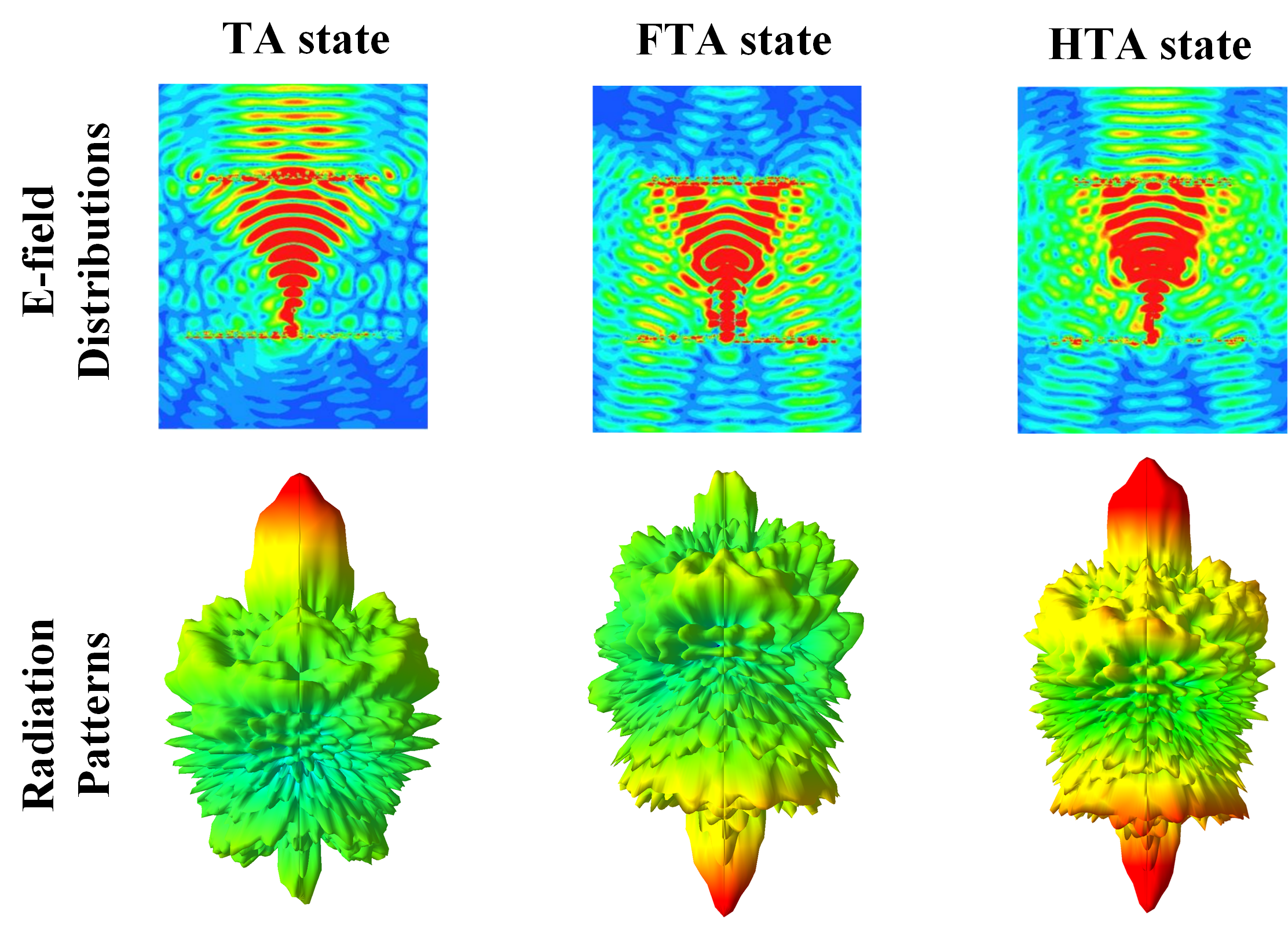}}
\caption{Simulated E-field distributions and radiation patterns for the three proposed working states.}
\label{fig18}
\end{figure}

\section{Simulation and Measurement Results}
\subsection{Antenna Prototyping}

To validate the antenna design strategy, a prototype with overall dimensions of 410 mm × 390 mm × 223 mm was fabricated and measured, as depicted in Fig. 19(a). Fig. 19(b) illustrates the complete assembly of the antenna. First, the metasurface and the polarization grids are etched on a dielectric substrate of F4B with a thickness of 2 mm and a dielectric constant of ${\varepsilon _r}$  = 2.65. Figs. 19(c) and (d) show the metal patterns of the inner polarization conversion layer of TA and FTA, respectively. All the substrates were assembled with plastic nuts and bolts. Using a similar process, the feed antenna was etched on a Rogers 4003 dielectric substrate, and the PIN diodes, capacitance, and inductance were soldered to the corresponding positions in the feed network, as shown in Fig. 19(e) and Fig. 19(f).

\begin{figure}[!t]
\centerline{\includegraphics[width=\columnwidth]{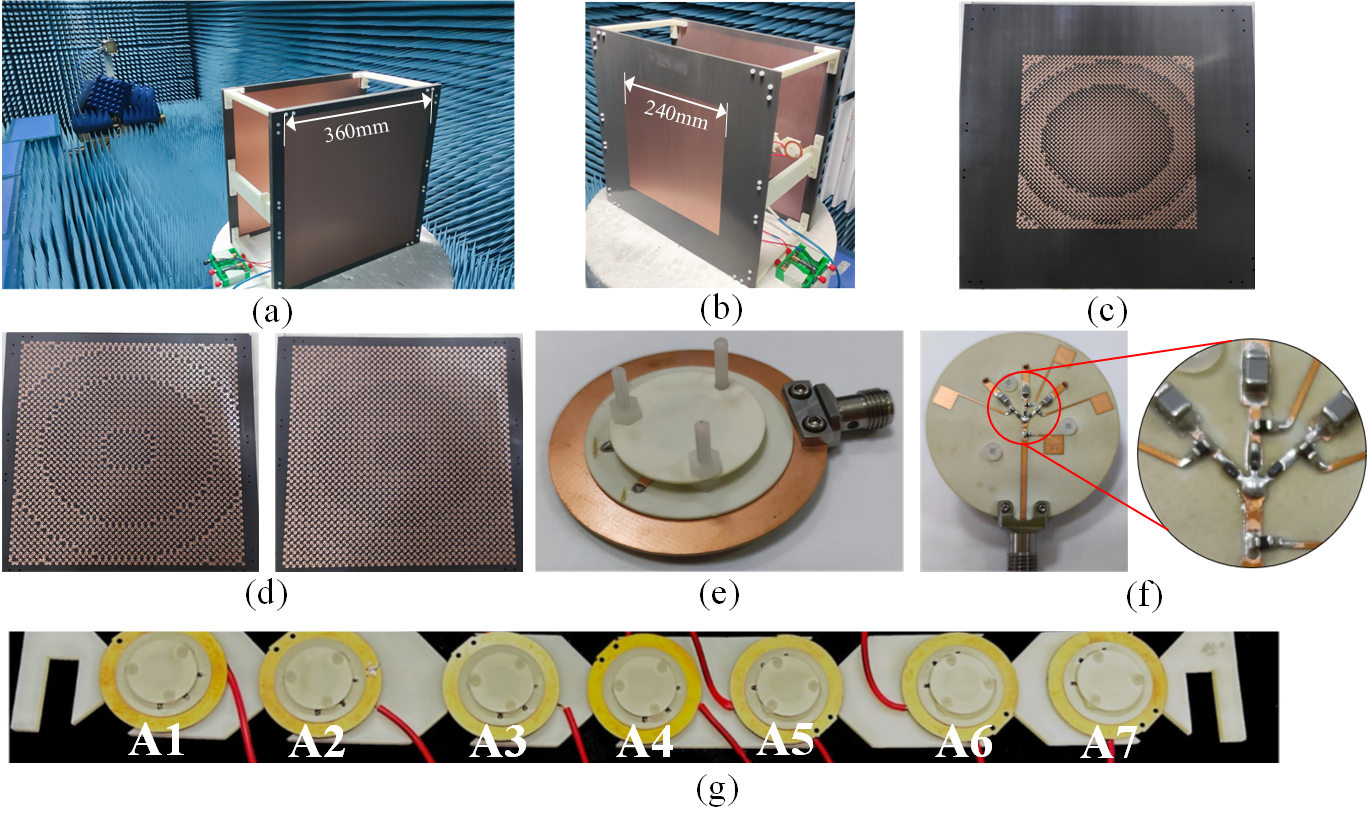}}
\caption{(a) Far-field measurement. (b) Assembled array. (c) Metasurface layer of TA. (d) Metasurface layer of FTA. (e) Antenna feed. (f) Polarization-switching feeding network. (g) Structure of MFA. }
\label{fig19}
\end{figure}

\begin{figure}[!t]
\centerline{\includegraphics[width=\columnwidth]{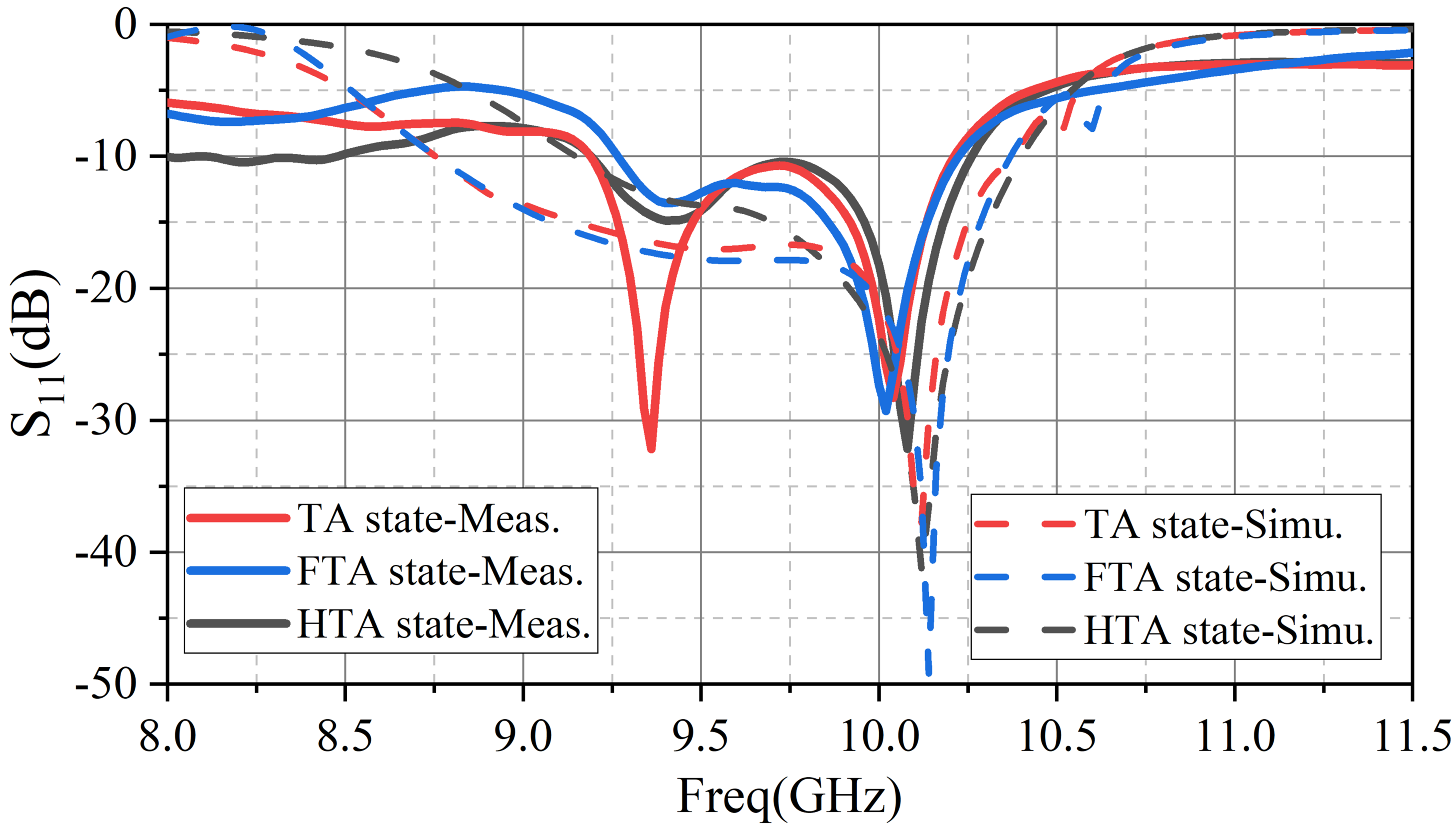}}
\caption{Simulated and measured ${{\rm{S}}_{11}}$ parameters of different states.}
\label{fig20}
\end{figure}

\begin{table*}[t]
	\centering
	\caption{Working States of The Antenna System}
	\label{tab:1}  
	\begin{tabular}{cccc ccc}
		\hline\hline\noalign{\smallskip}	
		\textbf{The Polarization of MFA} &\textbf{Working States} &\textbf{Feed Antenna Selection} &\textbf{Direction of Generated Beam} &\textbf{Polarization of Generated Beam}  \\
		\noalign{\smallskip}\hline\noalign{\smallskip}
		x-Polarization &TA State & A2-A6 &Upward &y-Polarization \\
		y-Polarization &FTA State & A1-A7 &Backward &y-Polarization \\
        45°Oblique-Polarization &HTA State & A1-A7 &Upward and Backward &y-Polarization \\
		\noalign{\smallskip}\hline
	\end{tabular}
\end{table*}

\begin{figure*}[!t]
\centerline{\includegraphics[width=0.9\textwidth]{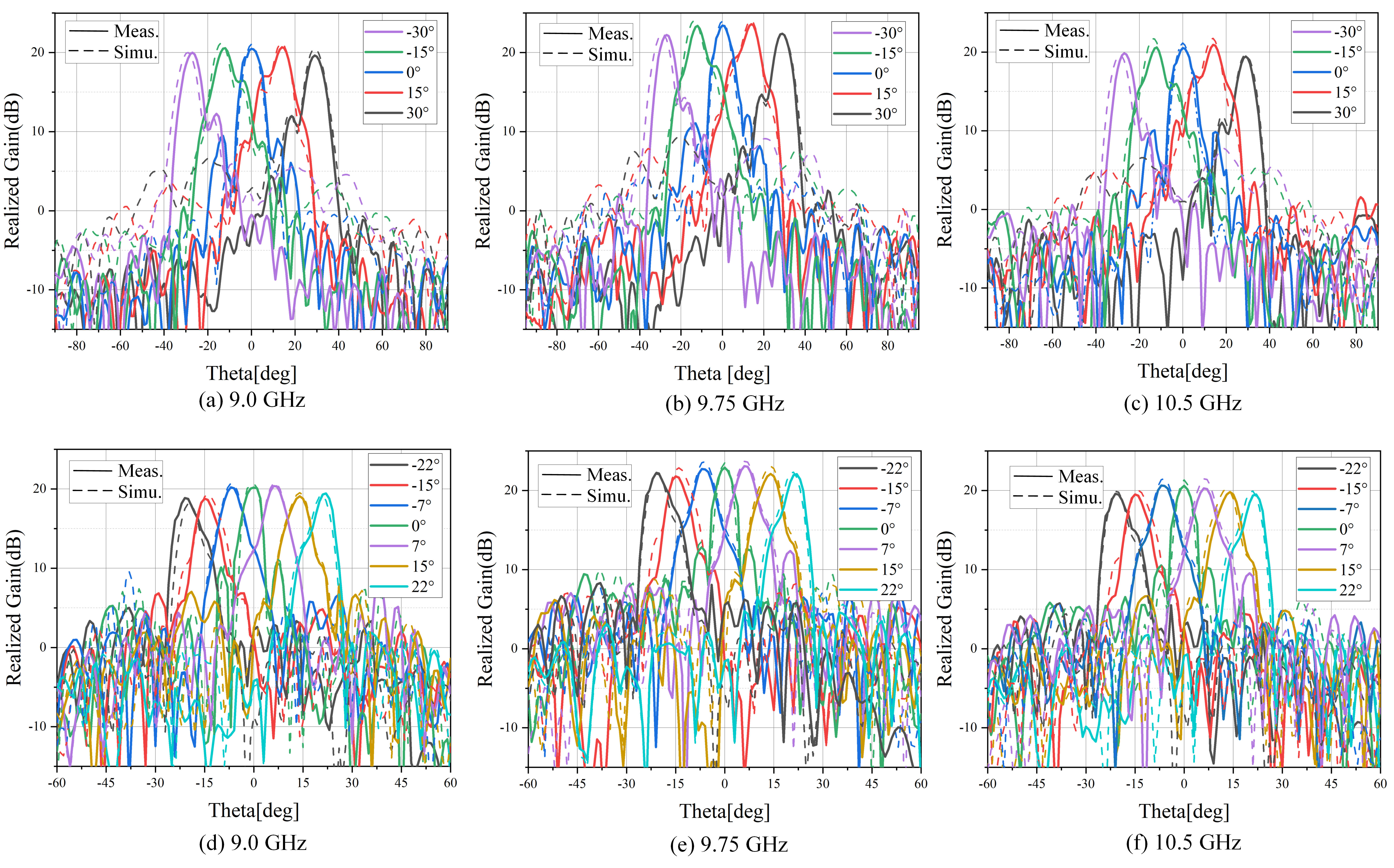}}
\caption{Radiation patterns of beam scanning in TA and FTA states. (a)-(c) are patterns at TA state, (d)-(f) are patterns at FTA state.}
\label{fig21}
\end{figure*}

\begin{figure*}[!t]
\centerline{\includegraphics[width=0.86\textwidth]{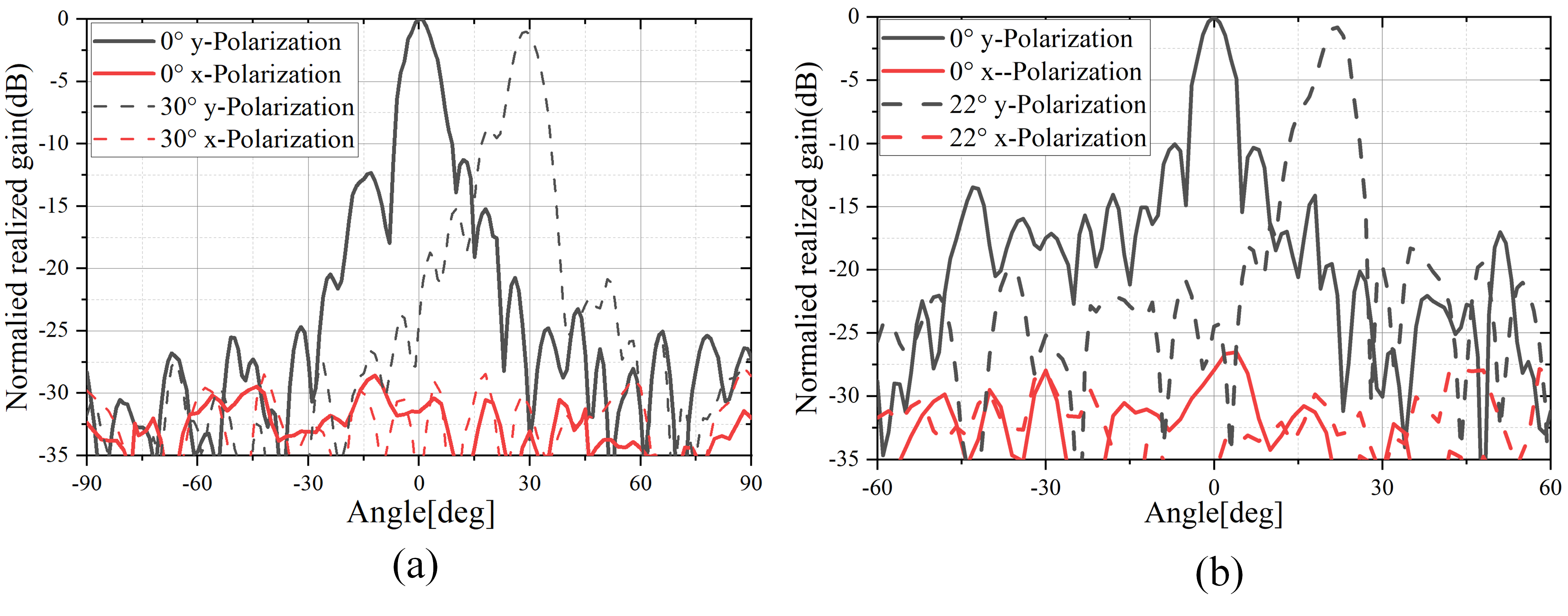}}
\caption{(a) Measured Co-polarization and Cross-polarization radiation patterns in 9.75 GHz of TA state. (b) Measured Co-polarization and Cross-polarization radiation patterns in 9.75 GHz of FTA state.}
\label{fig22}
\end{figure*}

\begin{figure*}[!t]
\centerline{\includegraphics[width=0.86\textwidth]{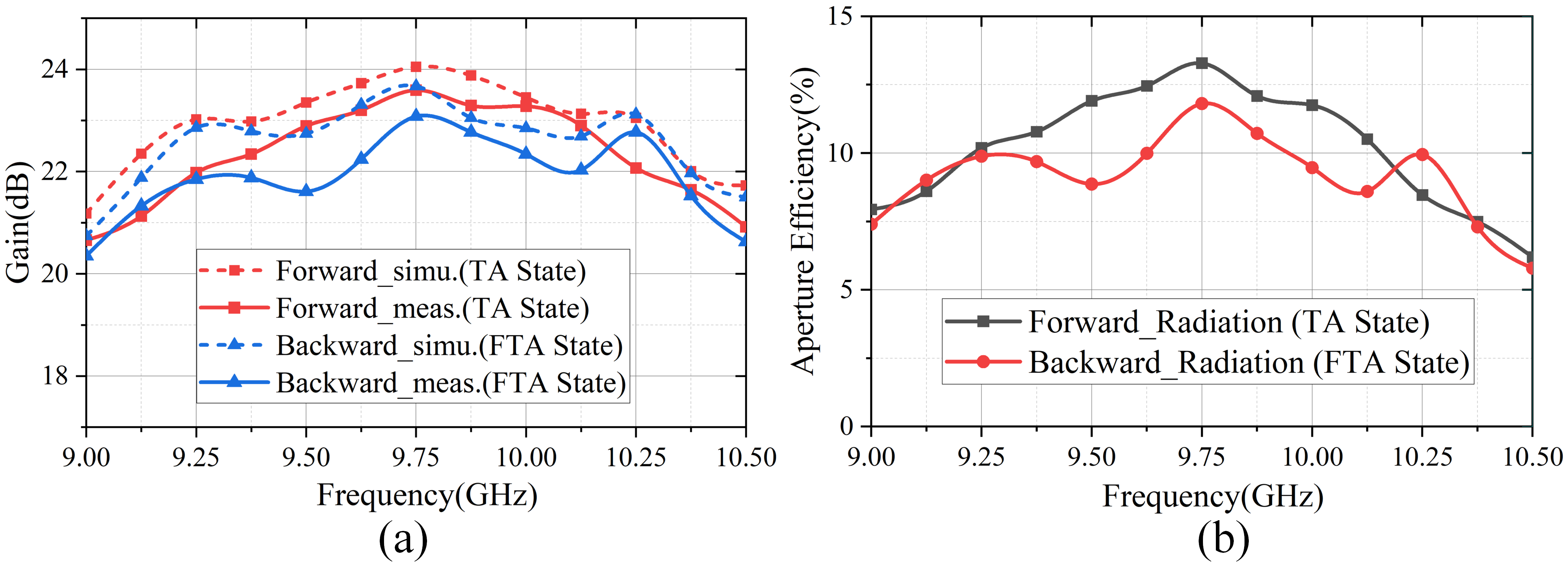}}
\caption{(a) Peak gains versus frequency for TA and FTA states. (b) Aperture efficiency versus frequency for TA and FTA states.}
\label{fig23}
\end{figure*}

\begin{figure*}[!t]
\centerline{\includegraphics[width=0.9\textwidth]{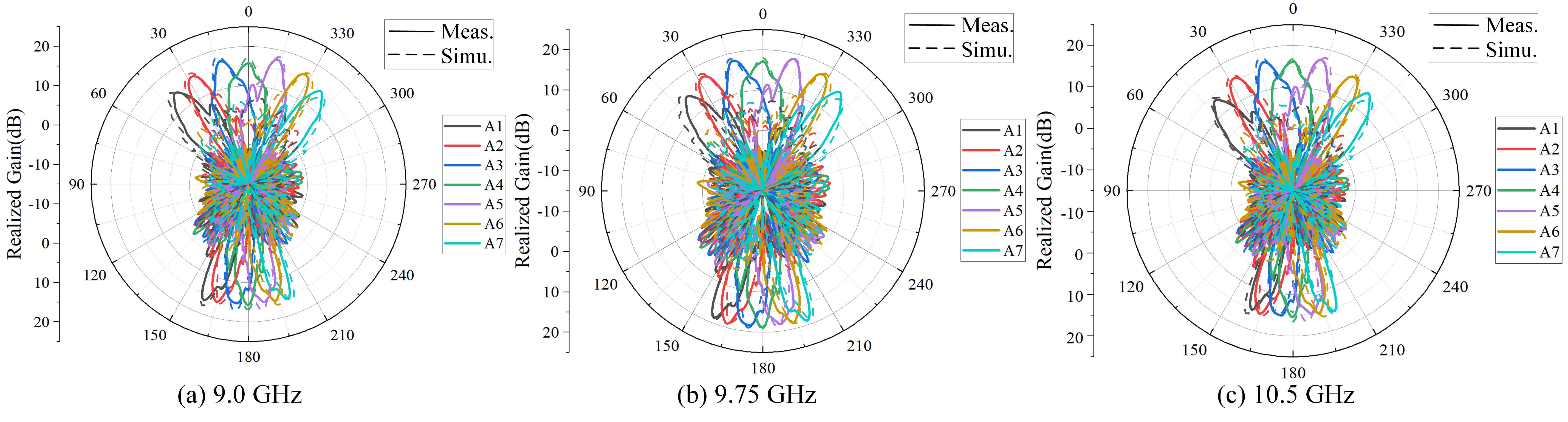}}
\caption{Radiation patterns of bidirectional beam scanning in HTA state. (a) 9 GHz. (b) 9.75 GHz. (c) 10.5 GHz.}
\label{fig24}
\end{figure*}

\begin{figure}[!t]
\centerline{\includegraphics[width=\columnwidth]{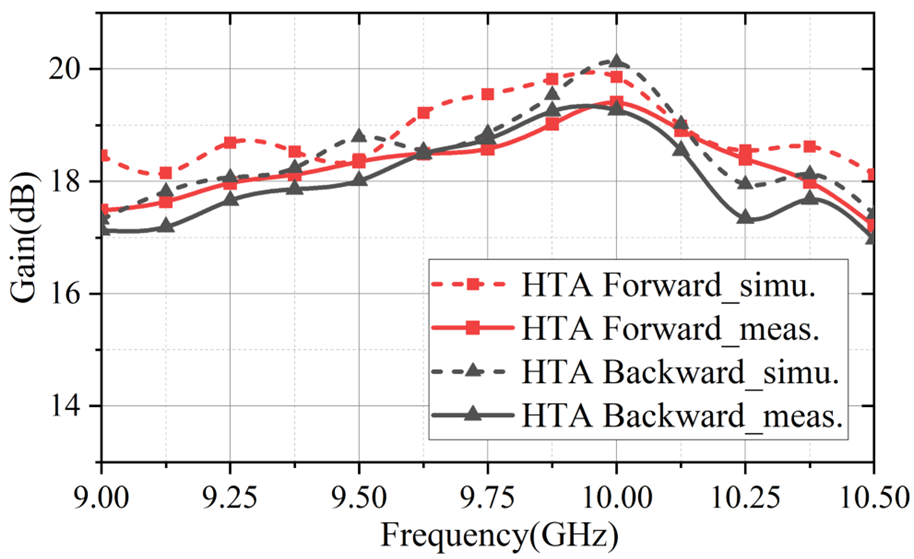}}
\caption{Peak gains versus frequency of forward and backward in HTA state.}
\label{fig25}
\end{figure}

\begin{figure}[!t]
\centerline{\includegraphics[width=\columnwidth]{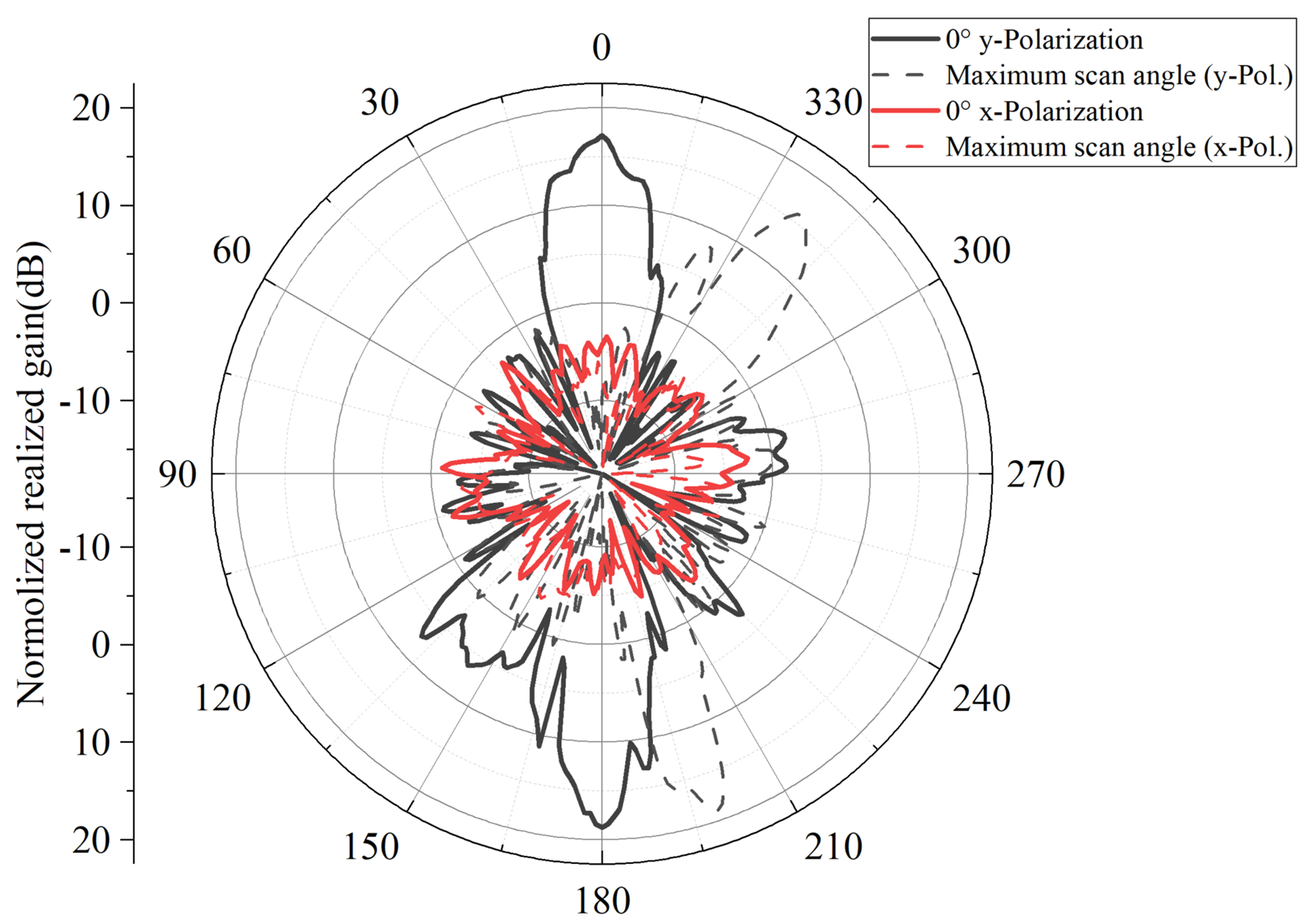}}
\caption{Measured Co-polarization and Cross-polarization radiation patterns of HTA state.}
\label{fig26}
\end{figure}

As shown in Fig. 19(g), the seven feeds of A1-A7 are non-uniformly arranged, and their lateral coordinates are +/-160 mm, +/-110 mm, +/-50 mm, and 0 mm. The feed lines of the MFA are connected to a single-pole multi-throw (SPMT) RF switch for beam switching, while the radiation states are selected by controlling the PIN diodes embedded in the feeding network. The far-field radiation patterns were measured in an anechoic chamber using a standard horn, as shown in Fig. 19(a). Table ${\rm I}$ lists the radiation characteristics of the three working states.

The simulated and measured ${{\rm{S}}_{11}}$ results are below -10 dB within the frequency range from 9.3 GHz to 10.3 GHz, as shown in Fig. 20. The feed antenna employs a parasitic patch to generate additional resonant points, thereby increasing the bandwidth. 

\subsection{TA and FTA Operation}
Fig. 21(a)-(e) shows the beam scanning patterns of TA and FTA at the frequency points of 9.0 GHz, 9.75 GHz, and 10.5 GHz, including measured and simulation results. When the MFA operates in the x-polarization (TA state), the A2 to A6 feeds are sequentially excited to produce forward beams with scanning angles of +/-30°, +/-15°, and 0°. As shown in Fig. 21(a)-(c), the peak gains for scanning angles of 0°, +/-15°, and +/-30° are 23.4 dBi, 23.6 dBi, and 22.4 dBi. The maximum scan loss of the beam in the TA state is  1.2 dB. When the MFA works in the y-polarization (FTA state), As shown in Fig. 21(d)-(f). The feeds A1 to A7 can be excited in turn to generate the main beams with scanning angles of +/-22°, +/-15°, +/-7°, and 0°. The peak gains for the scanning angles of 0°, +/-7°, +/-15°, and +/-30° are 22.9 dBi, 23.1 dBi, 22.1 dBi, and 22.2 dBi. The maximum scan loss of the beam is  1.0 dB. The discrepancy between the simulated and measured results is primarily attributed to the manufacturing error of the metasurface. Fig. 22 presents the measured co-polarization and cross-polarization radiation patterns. As shown in Fig. 22 (a), in the TA state, the SLLs of the beams are lower than -11.3 dB, and the cross-polarization level is lower than -28.5 dB. As the scanning angle increases, SLLs gradually rise, reaching a maximum of -8.0 dB. For the FTA state, the cross-polarization level is below -26.5 dB, as shown in Fig. 22(b). The peak SLLs are -10.1 dB at 0° and -16.4 dB at 22°.

Fig. 23 illustrates the gain and aperture efficiency for the TA and FTA states in the frequency range from 9.0 GHz to 10.5 GHz. As shown in Fig. 23(a), the maximum gain at 9.75 GHz is 23.6 dBi for the TA radiation state and 23.1 dBi for the FTA radiation state, and the 3 dB bandwidth is over 15.4$\%$ (9.0-10.5 GHz). The aperture efficiency for each radiating state is calculated using the maximum physical aperture (360 mm × 360 mm, shown in Fig. 19(a)), resulting in an aperture efficiency of 13.3$\%$ for the TA state, and 11.9$\%$ for the FTA state, respectively. It is worth noting that the enlarged bottom polarization grid layer in the upper TPRM is not included in the radiating aperture for the TA state. Therefore, a higher aperture efficiency of 29.9$\%$ can be achieved if the effective aperture size of the TA state (240 mm x 240 mm, shown in Fig. 19(b)) is used.

\begin{table*}[]
\centering
\caption{Comparison Of The Proposed Antenna And Related Works}
	\resizebox{\linewidth}{!}{\begin{tabular}{c|c|c|c|c|c|c|c|c}
		\hline	\hline
		Ref. &
		\begin{tabular}[c]{@{}c@{}}Frequency\\ (GHz)\end{tabular} &
		\begin{tabular}[c]{@{}c@{}}Gain\\ Bandwidth\end{tabular} &
		Beam-control method &
		Beam-scanning capability &
		Radiation states &
		\begin{tabular}[c]{@{}c@{}}Forward/\\ Backward gain (dBi)\end{tabular} &
        \begin{tabular}[c]{@{}c@{}}Aperture\\ efficiency ($\%$)\end{tabular} &
        \begin{tabular}[c]{@{}c@{}}Maximum scanning \\ angle (Gain loss)\end{tabular} \\ \hline
		{[}24{]} &
		5.35 &
		\begin{tabular}[c]{@{}c@{}}19.0\%\\ (1 dB)\end{tabular} &
		PIN diodes &
		Bidirectional &
		±z &
		17.2/15.4 &
        8.2/5.4 &
		\begin{tabular}[c]{@{}c@{}}±60°\\ (4.9 dB)\end{tabular} \\ \hline
		{[}26{]} &
		10 &
		\begin{tabular}[c]{@{}c@{}}14.0\%\\ (1 dB)\end{tabular} &
		N.A. &
		N.A. &
		\begin{tabular}[c]{@{}c@{}}+z, -z, \\ ±z\end{tabular} &
		25.0/25.5 &
        14/15 & 
		N.A. \\ \hline
		{[}28{]} &
		9.5 &
		\begin{tabular}[c]{@{}c@{}}19.5\%\\ (3 dB)\end{tabular} &
		PIN diodes &
		Bidirectional &
		\begin{tabular}[c]{@{}c@{}}+z, -z, \\ ±z\end{tabular} &
		21.4/21.0 &
        19.2/17.3 &
		\begin{tabular}[c]{@{}c@{}}±60°, ±45°\\ (\textgreater 5 dB, \textgreater 4 dB)\end{tabular} \\ \hline
		{[}31{]} &
		9.0 &
		\begin{tabular}[c]{@{}c@{}}6.7\%\\ (1 dB)\end{tabular} &
		N.A. &
		N.A. &
		+z, -z &
		18.5/13.1 &
		28/N.A. &
        N.A. \\ \hline
		{[}36{]} &
		26.0 &
		N.A. &
		RF T/R modules &
		\begin{tabular}[c]{@{}c@{}}2-D\\ Unidirectional\end{tabular} &
		+z &
		10.9/N.A. &
        N.A. &
		\begin{tabular}[c]{@{}c@{}}±45°\\ (6.7 dB)\end{tabular}  \\ \hline
		{[}38{]} &
		25.0 &
		N.A. &
		Mechanical control &
		Unidirectional &
		+z &
		30.0/N.A. &
        28.5/N.A. &
		\begin{tabular}[c]{@{}c@{}}±30°\\ (1.6 dB)\end{tabular} \\ \hline
		\textbf{\begin{tabular}[c]{@{}c@{}}This\\ work\end{tabular}} &
		\textbf{9.75} &
		\textbf{\begin{tabular}[c]{@{}c@{}}15.4\%\\ (3 dB)\end{tabular}} &
		\textbf{\begin{tabular}[c]{@{}c@{}}Switched\\ feeds\end{tabular}} &
		\textbf{Bidirectional} &
		\textbf{\begin{tabular}[c]{@{}c@{}}+z, -z, \\ ±z\end{tabular}} &
		\textbf{23.6/23.1} &
        \textbf{13.3/11.9} &
		\textbf{\begin{tabular}[c]{@{}c@{}}±30°, ±22°\\ (1.2 dB, 1.0 dB)\end{tabular}} \\ \hline \hline
	\end{tabular}}
\end{table*}

\subsection{HTA Operation}
As stated above, the HTA state is enabled by switching the polarization to ±45 degrees. By exciting A1-A7 in MFA sequentially, forward beams with scan angles of +/-40°, +/-30°, +/-15°, and 0° and backward beams with scan angles of +/-22°, +/-15°, +/-7°, and 0° can be simultaneously generated. Fig. 24(a)-(c) shows the beam scanning patterns of the HTA at the frequency points of 9.0 GHz, 9.75 GHz, and 10.5 GHz, respectively. Since the 45° oblique-polarized electromagnetic wave can be decomposed into x-polarization and y-polarization waves with equal amplitude and phase, the resulting forward scanning beam and backward scanning beam are only half of the energy of the original two states relative to the previous two states.

As shown in Fig. 25, at 10 GHz, the HTA state achieves peak gains of 19.4 dBi and 19.3 dBi in the forward and backward directions, respectively, with a 3 dB gain bandwidth range exceeding 15.4$\%$ (9.0 to 10.5 GHz). As illustrated in Fig. 26, for forward radiation, the SLLs at 0° and 40° are less than -11.7 dB and -8.0 dB, respectively. For backward radiation, the SLLs at 0° and 22° are less than -8.0 dB and -17.3 dB, respectively. The bidirectional cross-polarization level remains below -25.0 dB.

\subsection{Discussion}
Table ${\rm I}{\rm I}$ summarizes the performance of the proposed design compared to previous related works. Unlike prior research, our design facilitates flexible beam-switching between forward, backward, and bidirectional radiation states, avoiding the extensive use of PIN diodes in the antenna aperture and thereby reducing the complexity of system control. To achieve a simple and efficient design, we implemented a novel low-power tuning mechanism. This approach facilitates seamless switching between different operating states. Furthermore, the design maintains a reliable level of bidirectional radiation gain. It is noteworthy that, compared to designs using a horn as the feed for backward radiation [24], [26], [28], our design employs a folded transmitarray configuration. This setup causes some blockage due to the feed array, resulting in a slight decrease in aperture efficiency. Nevertheless, for forward radiation mode, our design performs at a comparable level to the referenced works. To achieve cost-effective beam control, an SPMT switch is employed in the beam-switching network, although it introduces some insertion loss. However, this trade-off allows our design to achieve minimal gain degradation at larger scanning angles, even with a slightly limited scanning range.

\section{Conclusion}
This paper presents an innovative design for a bidirectional beam-scanning hybrid transmitarray in the X-band based on a feed polarization switching mechanism. By switching different polarization states of the feed array, three different radiation states in the +/-z direction are realized using the bifocal phase design theory combined with the polarization-selective metasurface design. Moreover, beam scanning in the forward, backward, and bidirectional modes is accomplished by modifying the polarization state of the MFA. In the +z direction, the scanning angle can span +/-30°, with a peak gain of 23.6 dBi and a maximum scan loss of 1.2 dB. In the -z direction, the scanning angle range reaches +/-22° with a peak gain of 23.1 dBi and a scan loss of 1.0 dB. This design achieves compact and robust bidirectional beam coverage with high gain and flexible beam control. These combined features make it a promising candidate for Integrated Sensing and Communication (ISAC), next-generation wireless backhaul communication systems, microwave sensor networks, etc.

\begin{IEEEbiography}[{\includegraphics[width=1in,height=1.25in,clip,keepaspectratio]{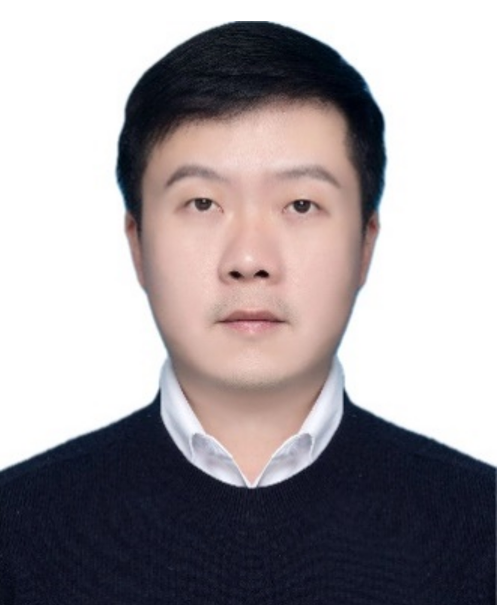}}]{Fan Qin} (Member, IEEE) received the B.S. degree in electronic information engineering and the Ph.D. degree in electromagnetic wave and microwave technology from Northwestern Polytechnical University, Xi’an, China, in 2010 and 2016, respectively. He was a visiting scholar with the University of Kent, Canterbury, UK, from 2013 to 2015. He joined the School of Telecommunications Engineering, Xidian University, Xi’an, China, in 2016. Currently, he is an Associate Professor with Xidian University. His research interests involve metamaterial antennas, millimeter wave antennas, vortex wave antennas, and metasurface-assisted wireless communication.
\end{IEEEbiography}

\begin{IEEEbiography}[{\includegraphics[width=1in,height=1.25in,clip,keepaspectratio]{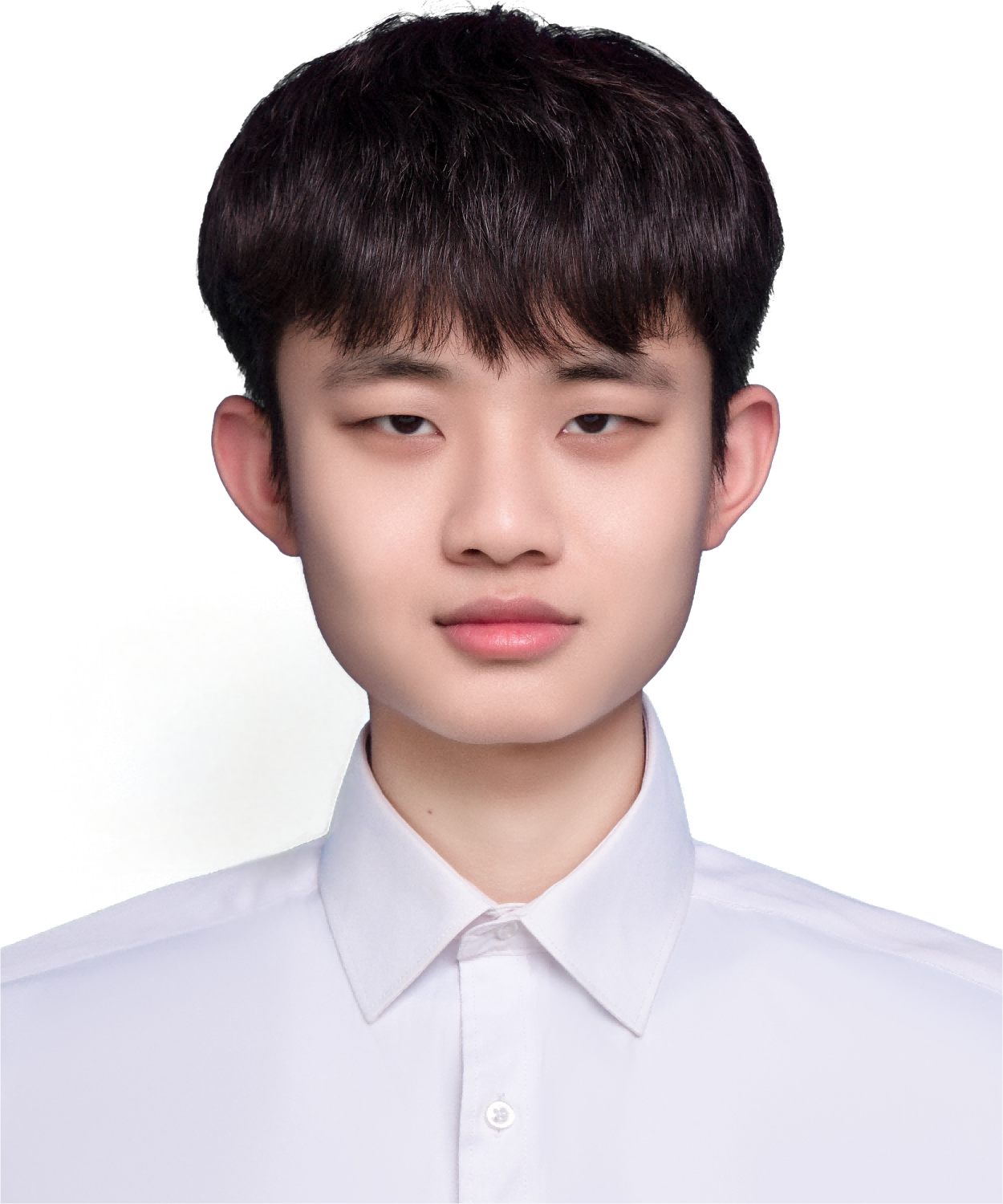}}]{Yifei Liu} (Student Member, IEEE) was born in Henan, China, in 2000. He received the B.S. degree in telecommunication engineering from Zhengzhou University, Zhengzhou, China, in 2022. 

He is currently a postgraduate student in information and telecommunication engineering at Xidian University, Xi'an, China. His current research interests involve metamaterial antennas and reconfigurable antennas. 
\end{IEEEbiography}

\begin{IEEEbiography}[{\includegraphics[width=1in,height=1.25in,clip,keepaspectratio]{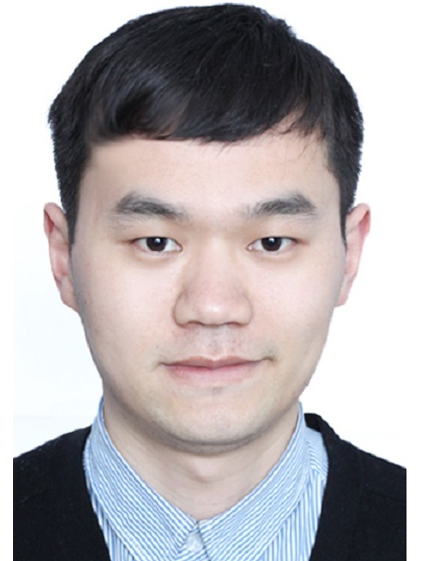}}]{Chao Gu} (Member, IEEE) received the B.Sc. and M.Sc. degrees from Xidian University, Xi’an, China, in 2009 and 2012, respectively, and the Ph.D. degree from the University of Kent, Canterbury, U.K., in 2017. From 2017 to 2018, he worked as a research associate at the University of Kent, where he led the development of additive manufacturing techniques for radio frequency components and antennas. Since August 2018, he has been a Senior Research Engineer at the Centre for Wireless Innovation, School of Electronics, Electrical Engineering and Computer Science, Queen’s University Belfast, U.K., actively participating in and leading extensive research and development tasks for various wireless applications, including space-based instrumentation and imaging, wireless power transfer, and sub-terahertz wireless solutions. His research interests include phased arrays, reconfigurable antennas, and metasurface antennas.
\end{IEEEbiography}

\begin{IEEEbiography}[{\includegraphics[width=1in,height=1.25in,clip,keepaspectratio]{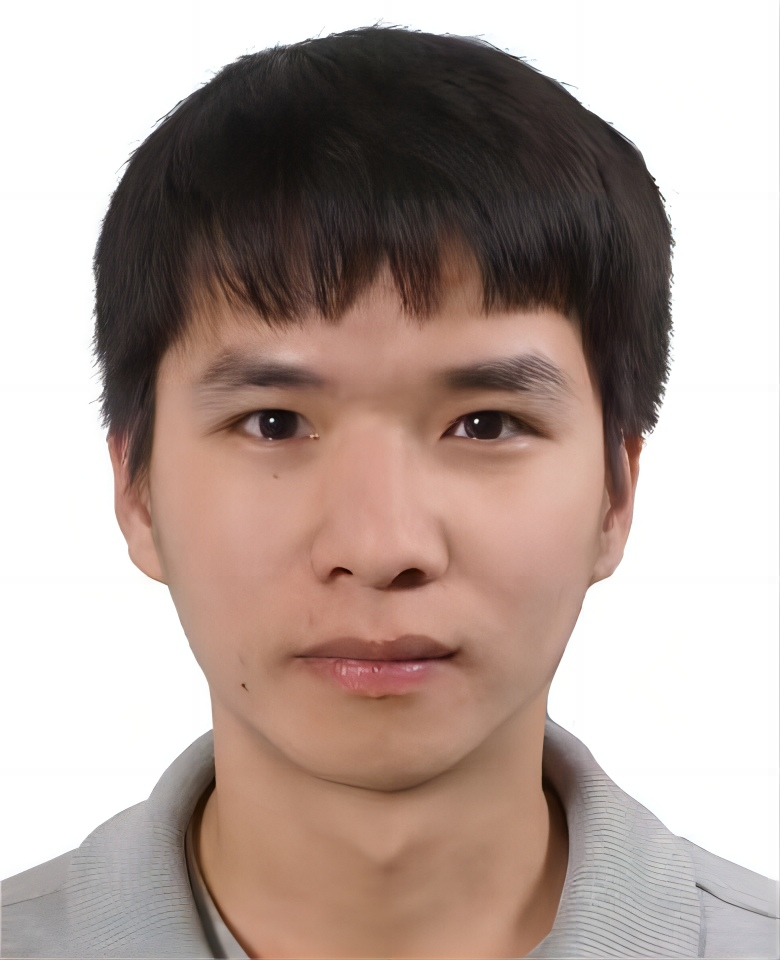}}]{Linfeng Zeng} (Student Member, IEEE) was born in Jiangxi, China. He received the B.S. degree in telecommunication engineering from Lanzhou University, Lanzhou, China, in 2020; and received the M.S. degree from the School of Telecommunication Engineering, Xidian University, Xi'an, China, in 2023. 

His current research interests include multibeam antennas and metasurface antennas.
\end{IEEEbiography}

\begin{IEEEbiography}[{\includegraphics[width=1in,height=1.25in,clip,keepaspectratio]{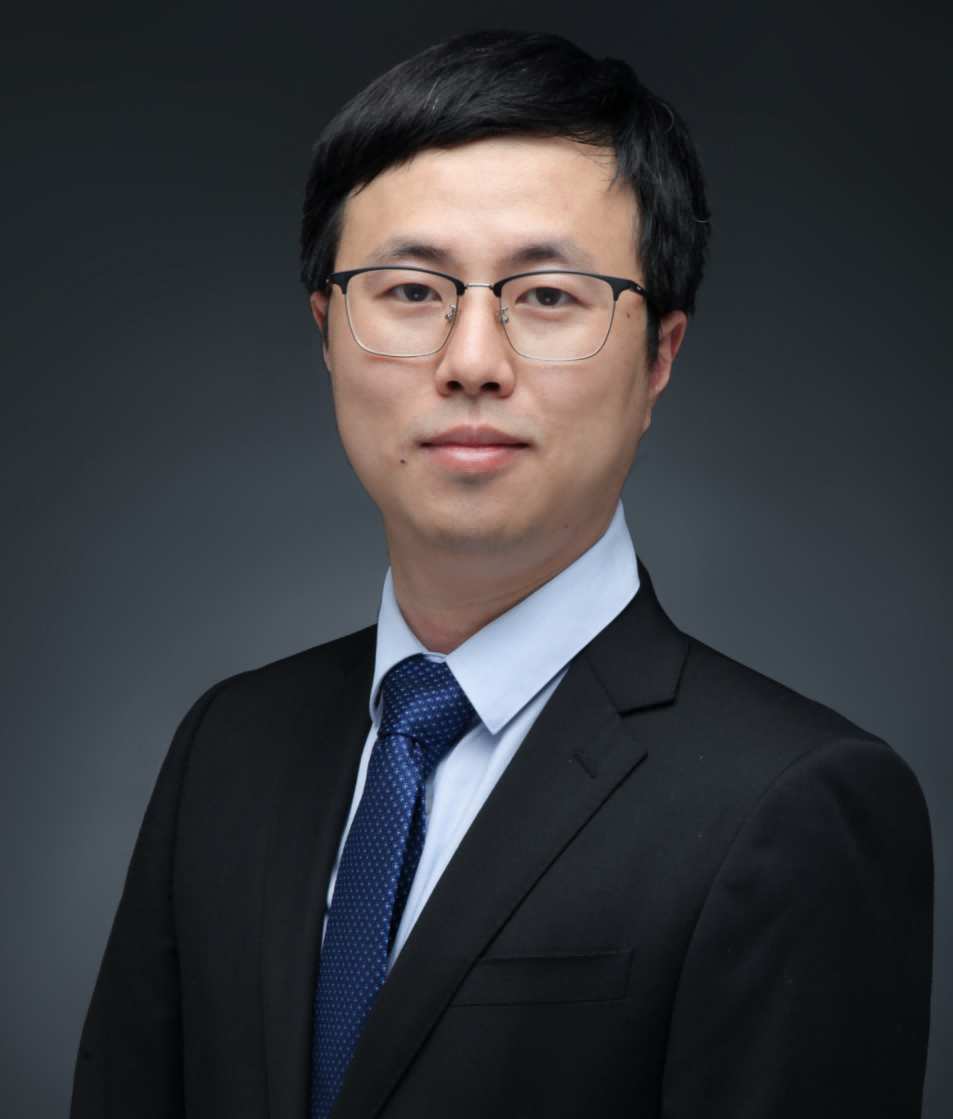}}]{Wenchi Cheng} (Senior Member, IEEE) received the B.S. and Ph.D. degrees in telecommunication engineering from Xidian University, Xian, China, in 2008 and 2013, respectively. He was a Visiting Scholar with the Department of Electrical and Computer Engineering, Texas A$\&$M University, College Station, TX, USA, from 2010 to 2011. He is currently a Full Professor with Xidian University. He has published more than 200 international journal and conference papers in \textit{IEEE Journal on Selected Areas in Communications}, \textit{IEEE Magazines}, \textit{IEEE TRANSACTIONS}, \textit{IEEE INFOCOM}, \textit{GLOBECOM}, and \textit{ICC}. His current research interests include 6G wireless networks, electromagnetic-based wireless communications, and emergency wireless information. He received the IEEE ComSoc Asia–Pacific Outstanding Young Researcher Award in 2021, the URSI Young Scientist Award in 2019, the Young Elite Scientist Award of CAST, and four IEEE journal/conference best papers. He has served or serving as the Wireless Communications Symposium Co-Chair for IEEE ICC 2022 and IEEE GLOBECOM 2020, the Publicity Chair for IEEE ICC 2019, the Next Generation Networks Symposium Chair for IEEE ICCC 2019, and the Workshop Chair for IEEE ICC 2019/IEEE GLOBECOM 2019/INFOCOM 2020 Workshop on Intelligent Wireless Emergency Communications Networks. He has served or serving as the ComSoc Representative for \textit{IEEE Public Safety Technology Initiative}, Editor for \textit{IEEE Transactions on Wireless Communications}, \textit{IEEE System Journal}, \textit{IEEE Communications Letters}, and \textit{IEEE Wireless Communications Letters}.
\end{IEEEbiography}

\begin{IEEEbiography}[{\includegraphics[width=1in,height=1.25in,clip,keepaspectratio]{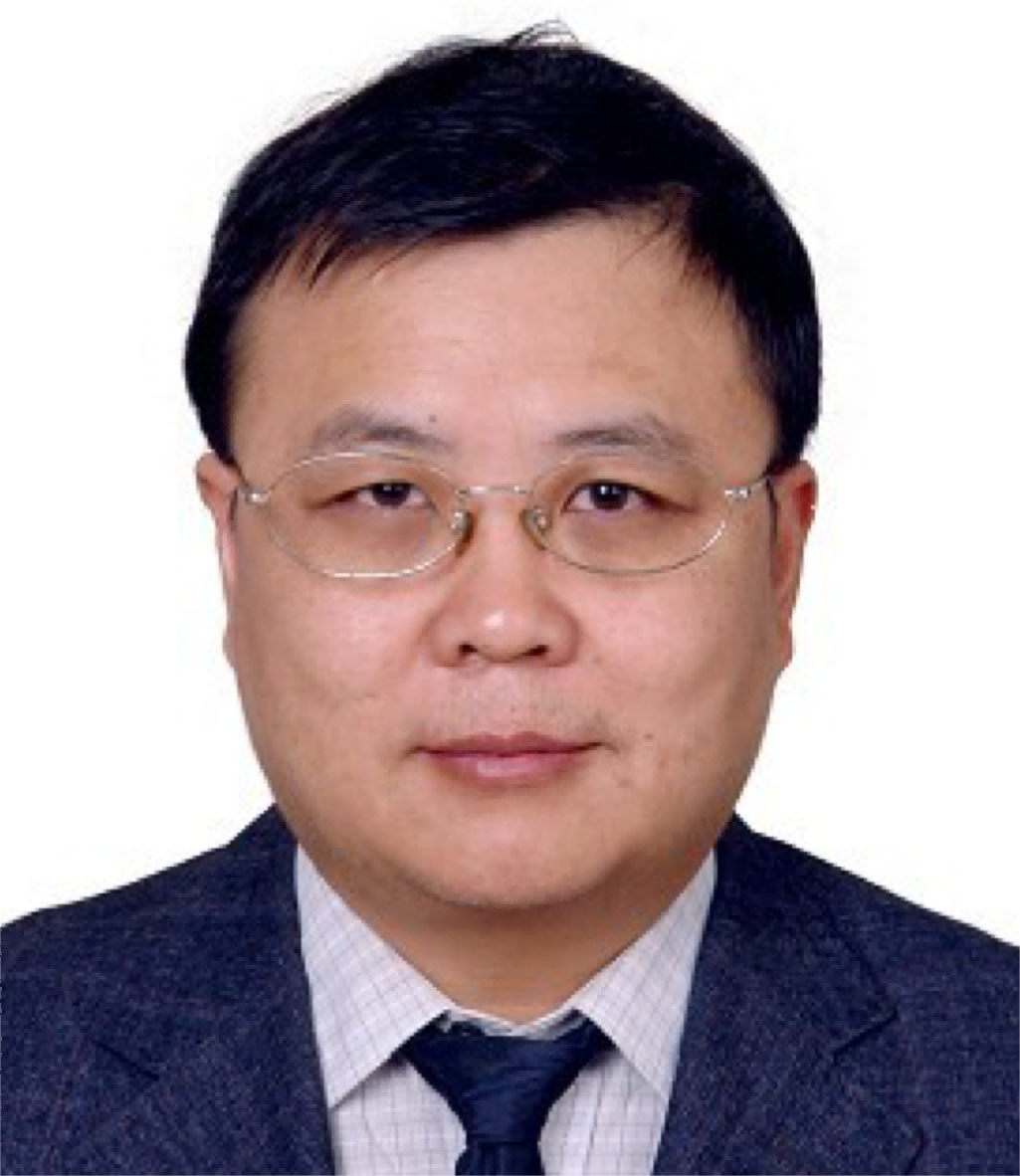}}]{Hailin Zhang} (Member, IEEE) received B.S. and M.S. degrees from Northwestern Polytechnic University, Xi’an, China, in 1985 and 1988 respectively, and the Ph.D. from Xidian University, Xi’an, China, in 1991. In 1991, he joined the School of Telecommunications Engineering, Xidian University, where he is a senior Professor of this school. He is also currently the Director of the Key Laboratory in Wireless Communications Sponsored by the China Ministry of Information Technology, a key member of the State Key Laboratory of Integrated Services Networks, one of the state government's specially compensated scientists and engineers, a field leader in Telecommunications and Information Systems in Xidian University, an Associate Director of National 111 Project. Dr. Zhang’s current research interests include key transmission technologies and standards on broadband wireless communications for 5G and 5G-beyond wireless access systems. He has published more than 150 papers in journals and conferences.
\end{IEEEbiography}

\begin{IEEEbiography}[{\includegraphics[width=1in,height=1.25in,clip,keepaspectratio]{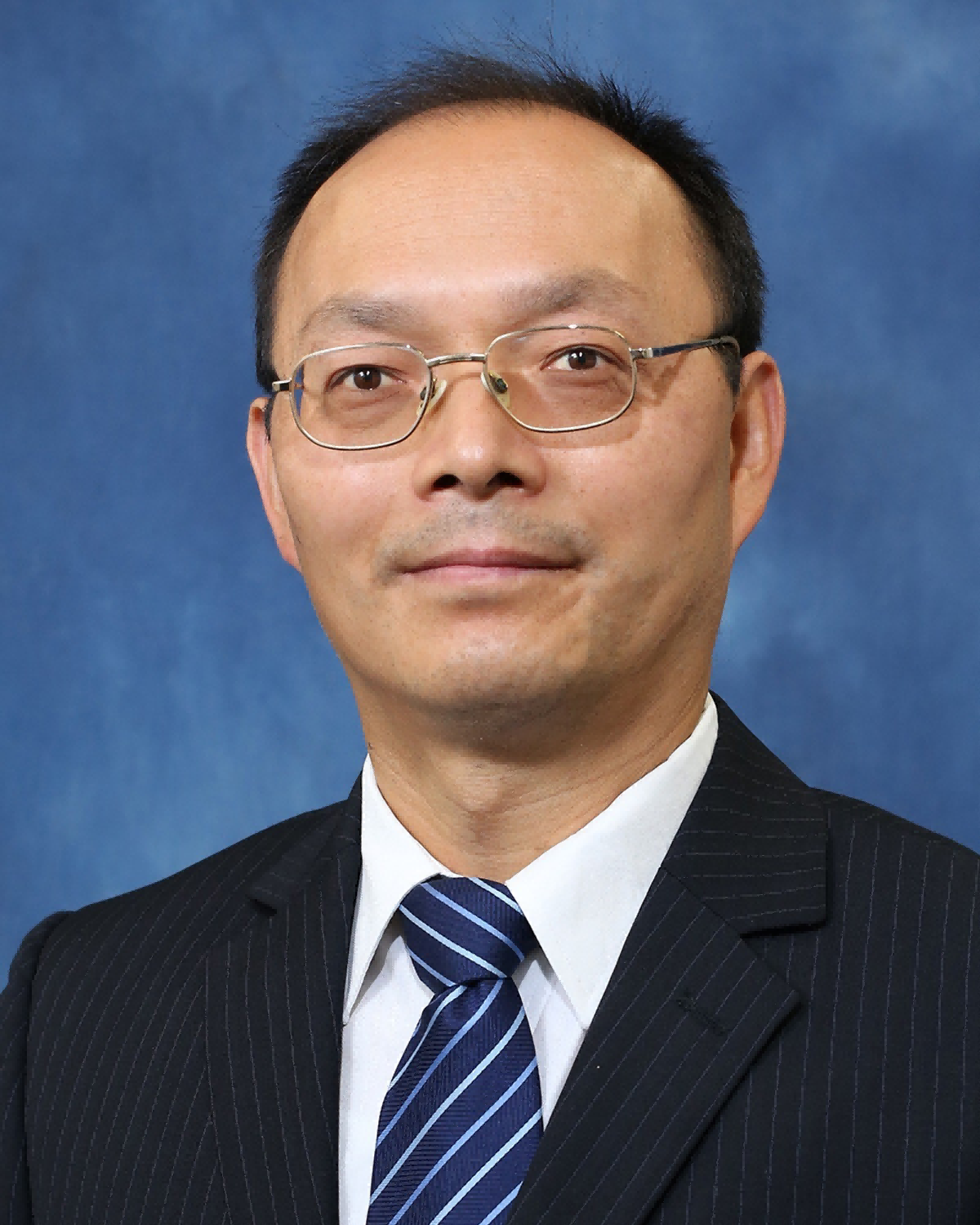}}]{Steven Gao} (Fellow, IEEE) received the PhD from Shanghai University, China. He is a Professor at the Department of Electronic Engineering at the Chinese University of Hong Kong, Hong Kong, where he is also the Director of the Center for Intelligent Electromagnetic Systems. Prior to joining CUHK, he was a Professor and Chair of RF/Microwave Engineering at the University of Kent (UKC), UK during 2013-2022, and became an Honorary Professor at UKC since 2022.

His research covers smart antennas, phased arrays, MIMO, reconfigurable antennas, broadband/multiband antennas, satellite antennas, and wireless systems (mobile and satellite communications, synthetic-aperture radars, IOT). He co-authored/co-edited 3 books (Space Antenna Handbook, Wiley, 2012; Circularly Polarized Antennas, IEEE $\&$ Wiley, 2014; Low-Cost Smart Antennas, Wiley, 2019), over 500 papers and 20 patents. 

He is a Fellow of IEEE and the Editor-in-Chief for \textit{IEEE Antennas and Wireless Propagation Letters}. He was a Distinguished Lecturer of IEEE Antennas and Propagation Society (2014-2016), and an Associate Editor for several international Journals (\textit{IEEE TAP}; \textit{Radio Science}; \textit{Electronics Letters}; \textit{IET Circuits}, \textit{Devices and Systems}, etc). He served as the Lead Guest Editor of \textit{Proceedings of the IEEE} for a Special Issue on “Small Satellites” (2018), and the Lead Guest Editor of \textit{IEEE Trans. on Antennas and Propagation} for Special Issues on "Low-Cost Wide-Angle Beam-Scanning Antennas” (2022) and "Antennas for Satellite Communication" (2015), and a Guest Editor of \textit{IET Circuits, Devices $\&$ Systems} for a Special Issue in “Photonic and RF Communications Systems” (2014). He was the UK’s Representative in the European Association on Antennas and Propagation (EurAAP) during 2021-2022, General Chair or General Co-Chair of international conferences (LAPC 2013, UCMMT 2021), and was an Invited/Keynote Speaker at many conferences. He is the TPC Chair of ISAPE 2024.
\end{IEEEbiography}
\end{document}